\pdfoutput=1
\documentclass[3p,12pt,nopreprintline,review,amsmath,amssymb,longbibliography,ofootinbib]{elsarticle}
\usepackage[activate={true,nocompatibility},final,tracking=true,kerning=true,spacing=true]{microtype}
\usepackage{epigraph}
\usepackage{mathrsfs}
\usepackage{graphicx}
\usepackage{siunitx}
\usepackage{seqsplit}
\usepackage{array}
\usepackage[T1,T2A]{fontenc}
\usepackage[utf8]{inputenc}
\usepackage{caption}
\usepackage{subcaption}
\usepackage{epstopdf}
\usepackage{numprint}
\usepackage{amsmath}
\usepackage{amssymb}
\usepackage{circuitikz}
\usepackage[colorlinks = true]{hyperref}

\newcommand\CapBlueLink{\href{http://datasheets.avx.com/AVX-SCC-LE.pdf}{AVX-SCCS20B505PRBLE}}
\newcommand\CapGreenLink{\href{https://www.eaton.com/content/dam/eaton/products/electronic-components/resources/data-sheet/eaton-hv-supercapacitors-cylindrical-cells-data-sheet.pdf}{Eaton-HV1020-2R7505-R}}
\newcommand\CapBlackLink{\href{https://www.mouser.com/datasheet/2/88/DCN_series-553005.pdf}{IC-505DCN2R7Q}}
\newcommand\CapWhiteLink{\href{https://www.mouser.com/datasheet/2/257/CE_2017_Datasheet_2_7V5F_3001974_EN_1-1274199.pdf}{Nesscap-ESHSR-0005C0-002R7}}
\newcommand\figureRef[2]{\hyperref[#1]{\ref*{#1}#2}}

\begin{document}
\title{On Supercapacitors Time--Domain Spectroscopy. \\
$C/R$ Characteristic Slope.
}
\def\surname{}

\author{Dmitry Valentinovich \surname{Agafonov}}
\ead{phti@lti-gti.ru}
\author{Arina Romanovna \surname{Kuznetsova}}
\ead{arinaspbgti@yandex.ru}
\address{St. Petersburg State Institute of Technology, Russia, 190013}
\author{Mikhail Evgenievich {Kompan}}
\ead{kompan@mail.ioffe.ru}
\author{Vladislav Gennadievich {Malyshkin}} 
\ead{malyshki@ton.ioffe.ru}
\address{Ioffe Institute, St. Petersburg, Russia, 194021}

\date{June 2, 2022}

\begin{abstract}
\begin{verbatim}
$Id: timedomain.tex,v 1.399 2024/02/11 10:19:40 mal Exp $
\end{verbatim}
A novel time-domain technique for supercapacitor characterization is developed, modeled numerically, and experimentally tested on a number of
commercial supercapacitors.
The method involves momentarily shorting a supercapacitor
for a brief duration, denoted as $\tau$,
and measuring first $\int Idt$ and second $\int I^2dt$
moments of current along with the potential before and after shorting.
The effective \hyperref[CCurrentC]{$C(\tau)$} and
\hyperref[RCurrentR:Dasha:Invariant]{$R(\tau)$} are then obtained
from charge preservation and energy dissipation invariants.
A linear behavior in $[R(\tau),C(\tau)]$ parametric plot
is observed by several orders of $\tau$.
This gives 
a $C/R$ characteristic slope:
how much $\Delta C$ we can ``gain'' if
we are ready to ``lose'' $\Delta R$ in internal resistance.
The $C/R$ characteristic slope characterizes
possible energy and power properties of the device
in terms of materials and technology used,
this is a measure of supercapacitor perfection.
The technique has been proven with experimental measurements
and then
 validated through computer
modeling, analytic analysis, and impedance spectroscopy
on a number of circuit types: transmission line, binary tree, etc.,
a new n-tree element (\hyperref[Zsol]{nTE}) is introduced.
The approach offers an alternative to low-frequency impedance spectroscopy
and methods outlined in the IEC 62391 standard.
It provides valuable insights into the performance and characteristics of supercapacitors.

\end{abstract}
\maketitle


\section{\label{intro}Introduction}

Distributed porous structure of supercapacitor
electrodes lead to an equivalent circuit in the form
of a distributed hierarchical $RC$ network that can be observed
in electrical measurements\cite{yoo2016fast,borenstein2017carbon,kompan2019reverse,il2020modeling,ghanbari2021self,pourkheirollah2023simplified}.
Supercapacitor measurement techniques can be classified as
frequency domain (impedance spectroscopy) and
time--domain (cyclic voltammetry, constant current charge/discharge regime\cite{maxwellSC,IEC62391}, etc.),
see
\cite{bard2022electrochemical,lasia2002electrochemical,bagotsky2015electrochemical}.
Multiple extensions to discharge techniques\cite{cheng2009assessments,yang2020comparative,allagui2018short,zhang2015supercapacitors,burke2011power,allagui2016spectral,baboo2023investigating} have been recently proposed.
In our previous work\cite{kompan2021inverse} we developed
a pulse-discharge type of measurement technique that allows to determine,
based on charge preservation invariant,
the capacitance $C(\tau)$ available at discharge time $\tau$.
In this work this approach has been extended and,
based on energy dissipation invariant,
the resistance $R(\tau)$ available at discharge time $\tau$
has been obtained. This $[R(\tau),C(\tau)]$ pair
allows us to build a time--domain analogue of impedance spectroscopy
$[\mathrm{Re} Z(\omega),\mathrm{Im} Z(\omega)]$.

Impedance spectroscopy is a powerful method to investigate properties of materials and electrodes reactions\cite{barsoukov2018impedance}.
It is a frequency domain
technique
where the system is probed with low amplitude AC harmonic signal,
the potential and current are measured with both amplitude and phase,
then complex impedance $Z=U/I$ is  obtained;
frequency range varies by many orders, typically  $10^{-3} \div 10^6~\mathrm{Hz}$,
what allows to obtain information on porous structures.
Interpretation is the most important step in impedance spectroscopy
application. An analysis consists in assuming
an equivalent circuit, then
element values are optimized to fit theoretical and experimental curves,
\href{https://www.scribner.com/software/68-general-electrochemistr376-zview-for-windows/}{ZView}
is a common tool.
Impedance spectroscopy application to supercapacitors characterization
has it's own specific. The supercapacitors equivalent circuit
is simpler than the ones of a general electrochemical system
and consists of a number of parallel and serial $RC$ branches
corresponding to porous structure of electrode materials.
An equivalent circuit of a porous system is a distributed $RC$ network;
the elements like constant phase element (CPE)
can be modeled with an infinite superposition of $RC$ branches\cite{valsa2013rc}.

\begin{figure}
\begin{tabular}{m{2em}m{4cm}m{3.7cm}m{3.7cm}}
(a)&
\begin{circuitikz}[line width=1pt,european resistors]
  \draw (0,0) node[anchor=east] {} to[short, *-] (0.5,0)
  to[R] ++(1.5,0) to[C] ++(1,0) to[short, -*] ++(0.5,0);
\end{circuitikz}
& \includegraphics[width=3.7cm]{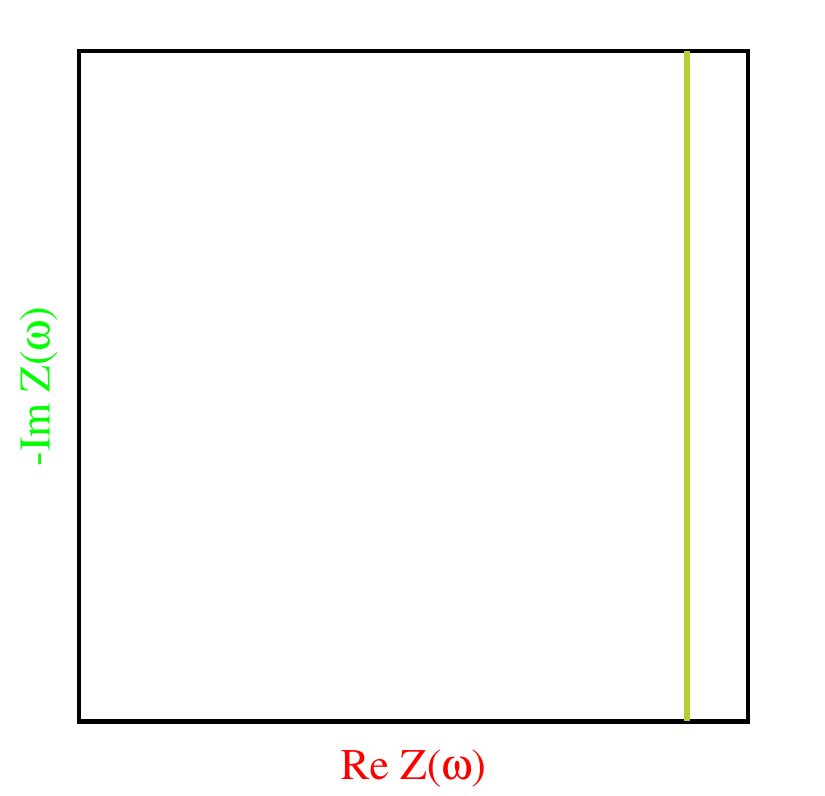}
& \includegraphics[width=3.7cm]{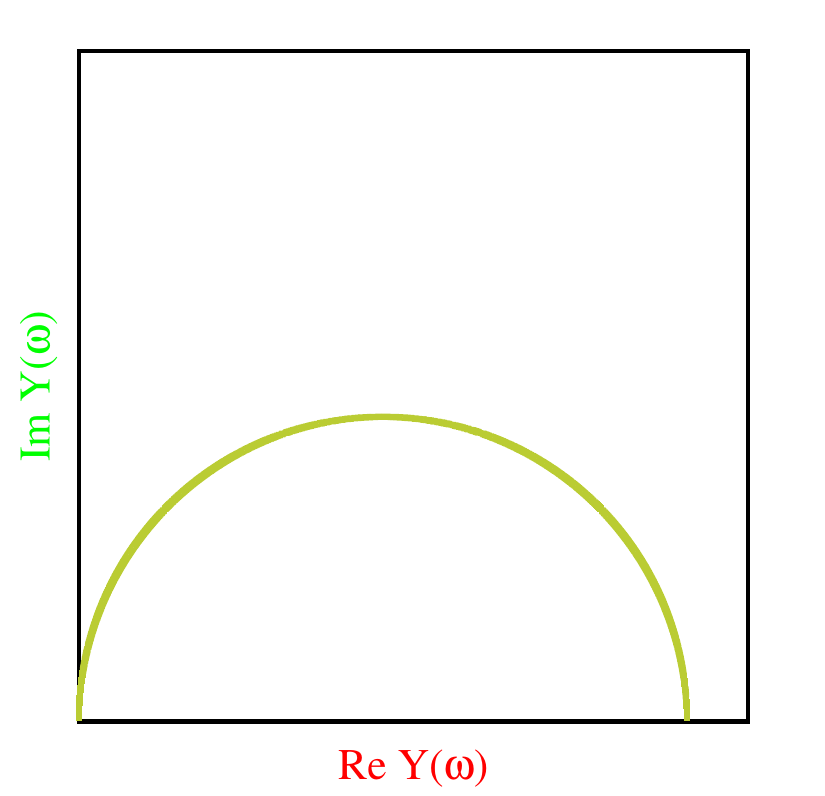}
\\
(b)&
\begin{circuitikz}[line width=1pt,european resistors]
  \draw (0,1) node[anchor=east] {} to[short, *-] ++(0.5,0)
         to[short, -] ++(0,0.5) 
  to[R] ++(2.5,0) to [short, -] ++(0,-0.5) to[short, -*] ++(0.5,0);
  \draw (0.5,1)  to[short, -] ++(0,-0.5)
  to[C] ++(2.5,0) to [short, -] ++(0,+0.5) ;
\end{circuitikz}
& \includegraphics[width=3.7cm]{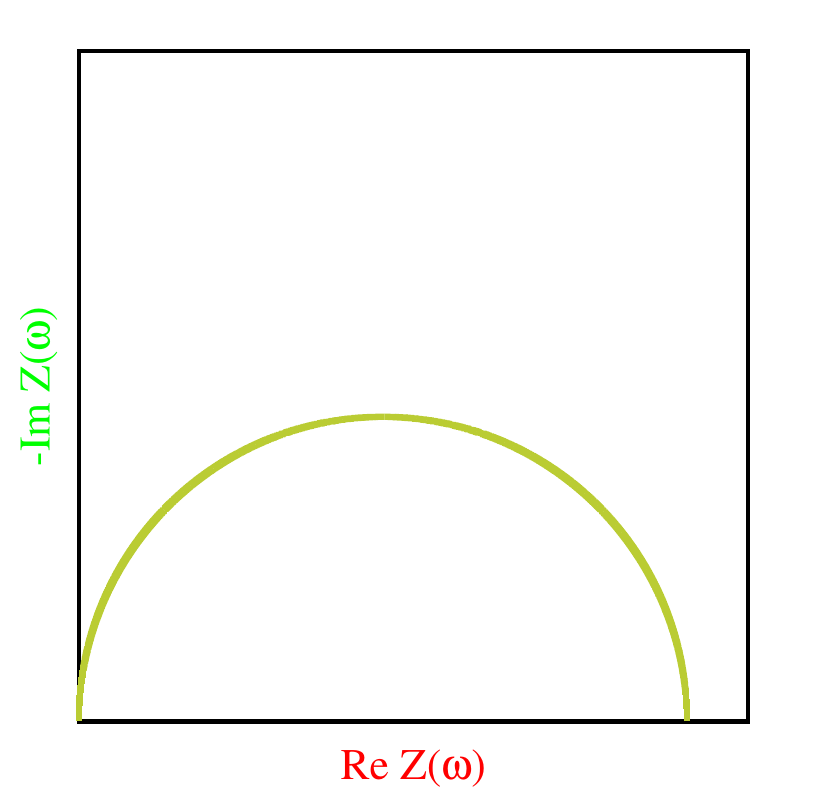}
& \includegraphics[width=3.7cm]{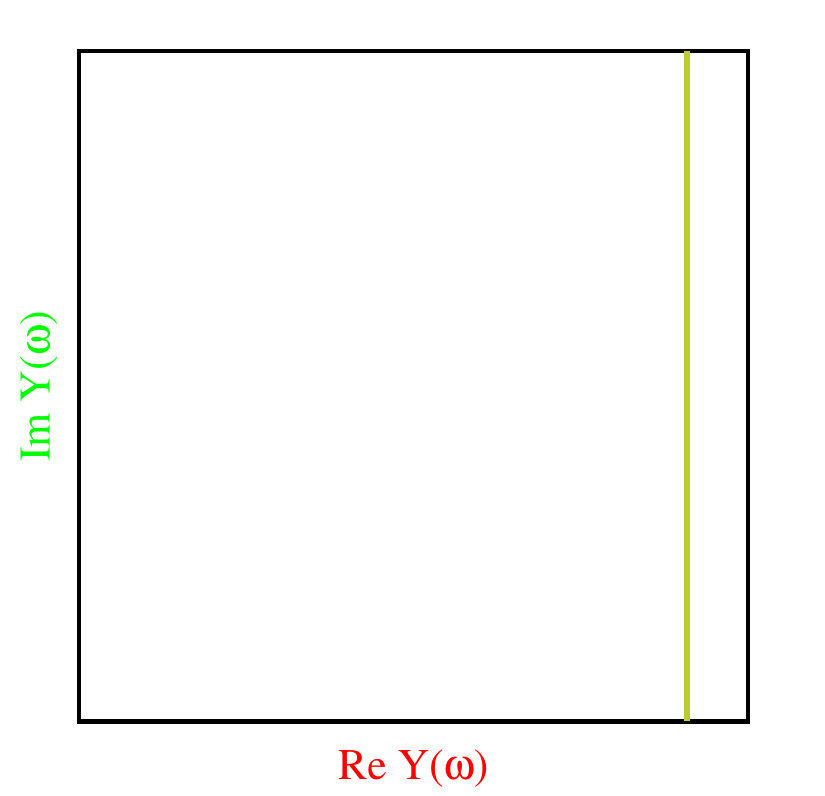}
\\
(c)&
\begin{circuitikz}[line width=1pt,european resistors]
  \draw (0,1) node[anchor=east] {} to[short, *-] ++(0.5,0)
         to[short, -] ++(0,0.5) 
  to[R=$R(U)$] ++(2.5,0) to [short, -] ++(0,-0.5) to[short, -*] ++(0.5,0);
  \draw (0.5,1)  to[short, -] ++(0,-0.5)
  to[C,l_=${C(U)=const;\omega=const}$] ++(2.5,0) to [short, -] ++(0,+0.5) ;
\end{circuitikz}
& \includegraphics[width=3.7cm]{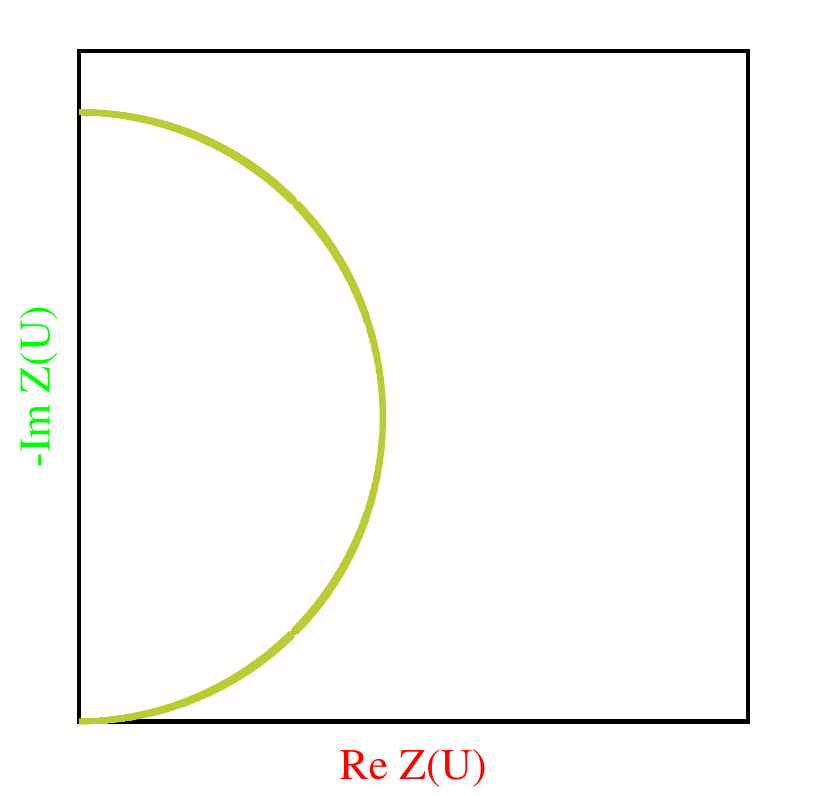}
& \includegraphics[width=3.7cm]{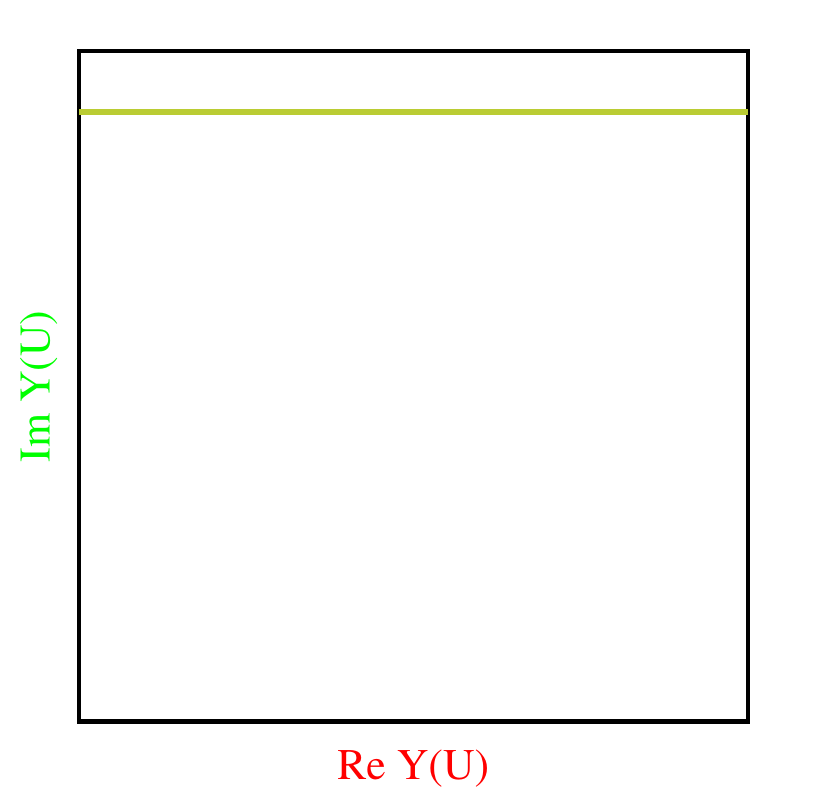}
\end{tabular}
\caption{
\label{differentRCcurcuits}
Several examples of impedance $Z$, admittance $Y=1/Z$,
and their conformal mapping.
}
\end{figure}

The most common tool for presenting impedance spectroscopy
theoretical and experimental results is
\href{https://en.wikipedia.org/wiki/Nyquist_stability_criterion#Nyquist_plot}{Nyquist plot},
which is a parametric plot of real and imaginary part of
impedance $Z$ with frequency $\omega$ as a parameter:
$[\mathrm{Re} Z(\omega),\mathrm{Im} Z(\omega)]$.
For simple supercapacitor models the $Z(\omega)$
is a vertical half-line for serially connected $C$ and $R$,
Fig. \figureRef{differentRCcurcuits}{a}
and a half-circle of Cole-Cole style
for parallel connection of  $C$ and $R$ (self-discharge),
Fig. \figureRef{differentRCcurcuits}{b}.
The systems with half-line and half-circle are related
to each other with the
conformal mapping $Y=1/Z$ ($Y$ is complex admittance, an inverse
to impedance $Z=1/Y$) that transforms\cite{lavrent1973methods,carrier2005functions} a half-line into a half-circle, i.e. an electrochemical system
with a vertical half-line (half-circle) in $[\mathrm{Re} Z(\omega),\mathrm{Im} Z(\omega)]$
would give a half-circle (vertical half-line) in $[\mathrm{Re} Y(\omega),\mathrm{Im} Y(\omega)]$.
This is a general result.
If the real/imaginary part of $Z$ (or $Y$) is a constant
and
the imaginary/real part depends \textsl{arbitrary} on a parameter $\xi$
(e.g. $\omega$, bias $U$, etc.) then a half-line in
$[\mathrm{Re} Z(\xi),\mathrm{Im} Z(\xi)]$
is transformed to a
half-circle in $[\mathrm{Re} Y(\xi),\mathrm{Im} Y(\xi)]$ and vice versa.
For example in
\cite{kompan2021impedanceHodograph}
we considered Fig. \figureRef{differentRCcurcuits}{c} system
to plot an impedance  $\left[\mathrm{Re}Z(U),\mathrm{Im}Z(U)\right]$
parametrically
where the impedance was measured at \textsl{fixed} frequency
with  voltage bias $U$ (varied parameter)
being applied to the system.
The bias
changed the value of $R(U)$, the value of $C(U)$
stayed constant. This corresponds to Fig. \figureRef{differentRCcurcuits}{c}
$[\mathrm{Re} Y(U),\mathrm{Im} Y(U)]$ plot. When converted to
$[\mathrm{Re} Z(U),\mathrm{Im} Z(U)]$ this
gives vertically oriented half-circle.

In addition to impedance/admittance type of transform $Y=1/Z$
it may be beneficial to consider other conformal mappings
such as $Y=1/(Z-R_1)$ for example to a system with parallel connection
of $R$ and $C$ with $R_1$ serially connected to them
\hbox{
\begin{circuitikz}[scale=0.4,transform shape,european resistors]
  \draw (0,1) node[anchor=east] {} to[short, *-] ++(0,0)
         to[R=$R_1$] ++(2.5,0)
         to[short, -] ++(0,0.5) 
  to[R] ++(2.5,0) to [short, -] ++(0,-0.5) to[short, -*] ++(0.5,0);
  \draw (2.5,1)  to[short, -] ++(0,-0.5)
  to[C] ++(2.5,0) to [short, -] ++(0,+0.5) ;
\end{circuitikz}
}.

Whereas classic impedance theory deals with 
$[\mathrm{Re} Z(\omega),\mathrm{Im} Z(\omega)]$ basis
(with conformal mapping possibly applied),
this form is not very convenient in supercapacitor
applications. First $\mathrm{Im} Z(\omega\to 0)$ diverges
at small $\omega$, second this basis
does not present directly how much energy/power
the supercapacitor can possibly generate during $\tau$.
In applications the most practical basis is capacitance vs internal resistance,
a parametric plot with characteristic time $\tau$ as a parameter;
this basis simultaneously characterizes energy and power properties of a SC.
For traditionally measured complex $Z(\omega)$
the $C(\tau)$ and $R(\tau)$ can be introduced as (\ref{CimpEq}) and (\ref{RimpEq})
respectively
with $\tau=1/\omega$ (\ref{TauRelOmega}),
this corresponds to $Z=R + \frac{1}{jwC}$.
A parametric plot $[R(\tau),C(\tau)]$ (dashed olive line)
is presented in Fig. \figureRef{cfigCircuitModellingCapacitance}{b}
for a model system
and in Fig. \ref{compareTW} for a real supercapacitor.
This $[R(\tau),C(\tau)]$ is a non-conformal mapping
of $[\mathrm{Re} Z(\omega),\mathrm{Im} Z(\omega)]$
with (\ref{CimpEq}), (\ref{RimpEq}).\footnote{
Compare with  $[\mathrm{Re} Y(\omega),\mathrm{Im} Y(\omega)]$
admittance
in Fig. \ref{differentRCcurcuits}
that can be obtained from impedance $[\mathrm{Re} Z(\omega),\mathrm{Im} Z(\omega)]$
with the
\href{https://mathworld.wolfram.com/ConformalMapping.html}{conformal mapping}
$Y=1/Z$.
}
The plot is very convenient.
The $R(\tau)$ and $C(\tau)$ increase together with $\tau$ 
 as the
electric current penetrates into deeper and
deeper pores of supercapacitor material.
A number of studies\cite{lozano2003influence,fuertes2004influence,fu2011hierarchical,wang2022controllable} show strong dependence of supercapacitor properties not only
on pores size, but also on their structure.
These dependencies show contributions of different pore sizes,
thus allow to predict the behavior at pulsed load\cite{danielyan2007increasing}
of different time length.
The low/high $\tau$ asymptotes give minimal/maximal internal resistance and capacitance.

All these results are obtained in frequency domain, solely from $Z(\omega)$ impedance data.
Impedance spectroscopy 
is a low current linear technique,
non--linear effects are problematic to study\cite{kompan2010nonlinear}.
At high AC amplitude the $Z(\omega)$
is measured in a non-linear regime what makes
it difficult to interpret the results.
The goal of this paper is to obtain the $[R(\tau),C(\tau)]$
from the measurements performed directly in \textsl{time domain}.
Since the developed technique is based on the measurement of
charge preservation and energy dissipation invariants --
obtained $[R(\tau),C(\tau)]$
remains valid even when working in a highly non-linear regime.
This allows us to take a completely new look into supercapacitor characterization.

\section{\label{timeDomain}Time Domain Spectroscopy:
Theory and Modeling}
In our previous work\cite{kompan2021inverse} a new technique
for supercapacitors characterization,
the inverse relaxation,
has been developed.
The technique consists in shorting a supercapacitor for a short duration $\tau$,
then switching it to the open circuit regime and measuring an initial
rebound and long-time relaxation. The current is measured
during shorting stage. The main result of the previous work was to obtain
the $C(\tau)$ that characterizes the distributiveness
of $RC$ network. In this work the technique has been extended
to obtain the $R(\tau)$.

\begin{figure}
\begin{tabular}{m{8cm}m{6cm}}
\noindent
\begin{circuitikz}[american voltages, european resistors]
\draw (7,-0.2) node[label={(a)}] {};
  \draw (0,2) node[anchor=east] {} to[short, *-] (0,2)
  to[R=$R_1$] (2,2)
  to[R=$R_2$] (4,2) ;
  \draw (4,2) [dashed] to [R=$R_3$] (6,2);
  \draw (6,2) [dashed]  to  (7,2);
  \draw (0,0) node[anchor=east] {} to[short, *-] (6,0);
  \draw (6,0) [dashed] to (7,0);
  \draw (2,2) to[C=$C_1$] (2,0);
  \draw (4,2) to[C=$C_2$] (4,0);
  \draw (6,2) [dashed] to[C=$C_3$] (6,0);
  \draw (0,2) to [open, v=SC] (0,0);
\end{circuitikz}

\bigskip
\begin{circuitikz}[american voltages, european resistors]
\draw (7,-0.2) node[label={(b)}] {};
  \draw (0,3) node[anchor=east] {} to[short, *-] (0,3) 
     to [short] (3,3);
  \draw (0,3) to [open, v=SC] (0,0)
   to [short, *-] (3,0) ;
 \draw (3,0) [dashed] to (7,0); 
 \draw (3,3) [dashed] to (7,3); 
 \draw (1,3) to[R=$R_1$] (1,1.3) to[C=$C_1$] (1,0);
 \draw (3,3) to[R=$R_2$] (3,1.3) to[C=$C_2$] (3,0);
 \draw (5,3) [dashed] to[R=$R_3$] (5,1.3) [dashed] to[C=$C_3$] (5,0);
\end{circuitikz}

\bigskip
\begin{circuitikz}[american voltages, european resistors ]
\draw (6,-4) node[label={(c)}] {};
\ctikzset{resistors/scale=0.75,capacitors/scale=0.4}
\draw (-0.5,0) node[anchor=east] {} to[short, *-] ++(0.1,0) to [R] ++(1.7,0) ;
\draw (-0.5,-2) node[anchor=east] {} to[short, *-] (-0.5,-2)
 to (0,-2) node[ground]{}; 
\draw (-0.5,0) to [open, v=SC] (-0.5,-2);
\draw (-0.5+0.1+1.7,0) to [R] (2,-2);
\draw (2,-2) to[short] ++(0,-0.2) to [C] ++(0,-0.1) to node[ground,scale=0.8]{} ++(0,0) ;
\draw (-0.5+0.1+1.7,0)  to [R] ++(0.42,1.8) to [short]   (2,3);
\draw (2,3) to[short] ++(0,-0.8) to [C] ++(0,-0.1) to node[ground,scale=0.8]{} ++(0,0) ;

\draw (-0.5+0.1+1.7,0) to[short] ++(0.5,0) to [C] ++(0.4,0) to node[ground,scale=0.8,rotate=90]{} ++(0,0) ;

\def\fbtree(#1,#2){
\def\xinit{#1}
\def\yinit{#2}
\draw (\xinit+2,\yinit+3) to [R] ++(1.5,1) to[short] ++(0,-0.2) to [C] ++(0,-0.1) to node[ground,scale=0.8]{} ++(0,0) ;
\draw (\xinit+2,\yinit+3) to [R] ++(1.5,-1) to[short] ++(0,-0.2) to [C] ++(0,-0.1) to node[ground,scale=0.8]{} ++(0,0);
\draw (\xinit+3.5,\yinit+4) to [R] ++(1.5,1.1) to[short] ++(0,-0.2) to [C] ++(0,-0.1) to node[ground,scale=0.8]{} ++(0,0);
\draw (\xinit+3.5+1.5,\yinit+4+1.1) [dashed] to[short] ++ (0.5,0.6);
\draw (\xinit+3.5+1.5,\yinit+4+1.1) [dashed] to[short] ++ (0.5,-0.2);

\draw (\xinit+3.5,\yinit+4) to [R] ++(1.5,-0.2) to[short] ++(0,-0.1) to [C] ++(0,-0.2) to node[ground,scale=0.8]{} ++(0,0);
\draw (\xinit+3.5+1.5,\yinit+4-0.2) [dashed] to[short] ++ (0.5,0.6);
\draw (\xinit+3.5+1.5,\yinit+4-0.2) [dashed] to[short] ++ (0.5,-0.2);

\draw (\xinit+3.5,\yinit+2) to [R] ++(1.5,0.8) to[short] ++(0,-0.2) to [C] ++(0,-0.1) to node[ground,scale=0.8]{} ++(0,0);
\draw (\xinit+3.5,\yinit+2) to [R] ++(1.5,-0.6) to[short] ++(0,-0.2) to [C] ++(0,-0.1) to node[ground,scale=0.8]{} ++(0,0);
\draw (\xinit+3.5+1.5,\yinit+2+0.8) [dashed] to[short] ++ (0.5,0.6);
\draw (\xinit+3.5+1.5,\yinit+2+0.8) [dashed] to[short] ++ (0.5,-0.2);
\draw (\xinit+3.5+1.5,\yinit+2-0.6) [dashed] to[short] ++ (0.5,0.6);
\draw (\xinit+3.5+1.5,\yinit+2-0.6) [dashed] to[short] ++ (0.5,-0.2);
}
\fbtree(0,0)
\fbtree(0,-5)

\end{circuitikz}

&
\noindent
\begin{circuitikz}[scale=1,line width=1pt, european resistors]
\draw (5,0) node[label={(d)}] {};
  \draw (0,4) node[anchor=east] {$U(t)$} to[short, *-] ++(3,0)
  to[nos=$\mathrm{Charge}$] ++(2,0)
  to[short, -*] ++(0,0) 
  to node[anchor=west] {$U_0$} ++(0,0) ;
  \draw (3,4)
  to[nos=$\mathrm{Short}$] ++(0,-2)
  to [R=$R_s$] ++(0,-1.5)
  to[short, -] ++(0,-0.5) ;
  \draw (0,0) to[short, *-*] ++(5,0);
  \draw (2.5,2.2)  node[anchor=east] {$U^*(t)$}
  to[short, *-] ++(0.5,0);
  \draw (1,0) to[C={SC}] ++(0,3)
   to[short, -] ++(0,1);
\end{circuitikz}
\bigskip
\hphantom{bigskip}
\bigskip
  \includegraphics[width=7cm]{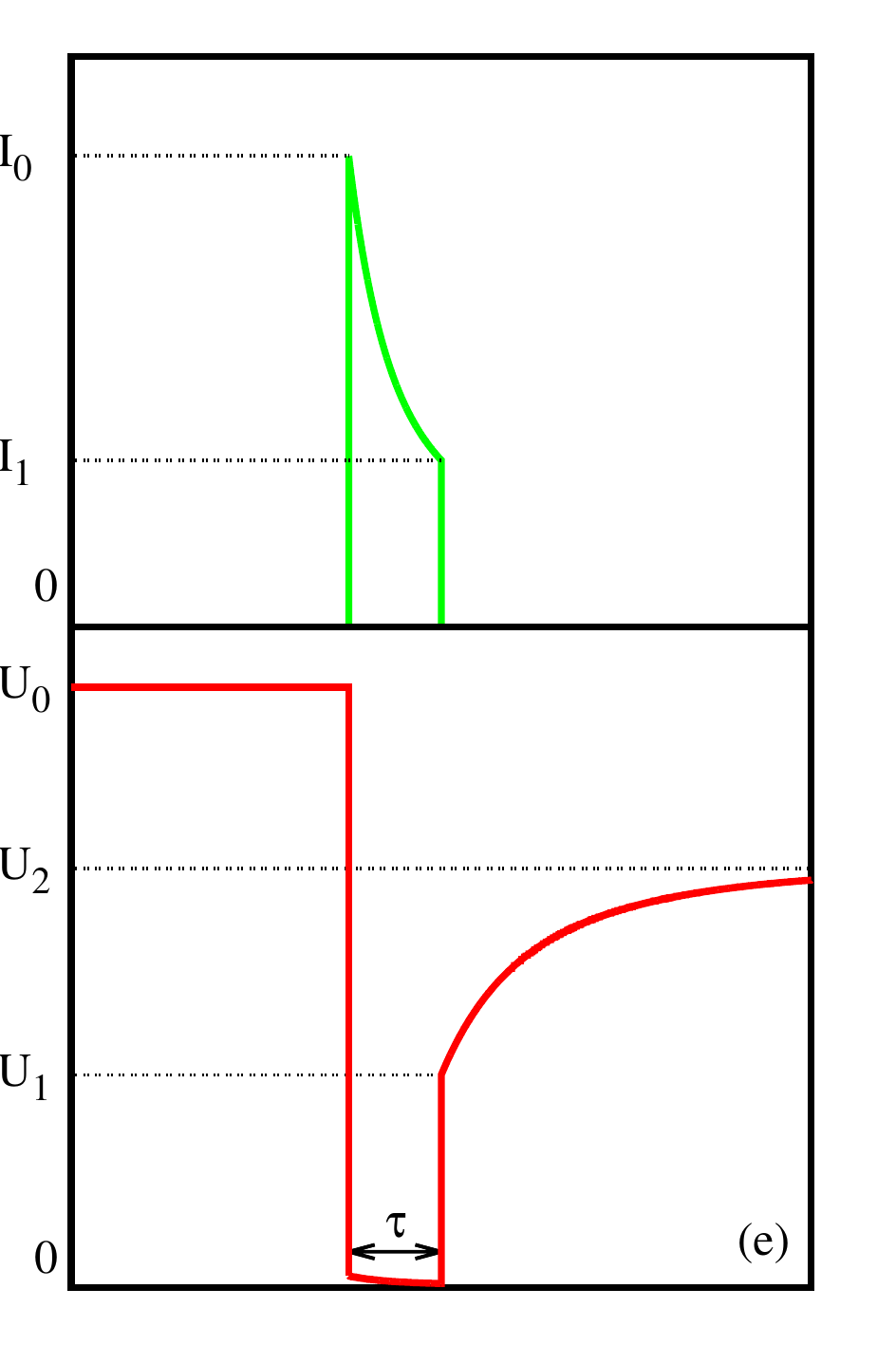}
\end{tabular}
  \caption{\label{cfigCircuit}
  Supercapacitor equivalent circuit models:
  transmission line (a),
  superposition of $RC$ (b) and
   multi--branch;  binary tree (c)
   is an example of a multi--branch model.
  Measurement circuit (d) and $U$, $I$  in time-domain (e).
  }
\end{figure}

For a distributed $RC$ network 
the $I(t),U(t)$ are measured in time--domain
as presented in Fig. \figureRef{cfigCircuit}{e}.
The main advantage of this time-domain measurement technique over
impedance spectroscopy is that the measurements
are performed in high-current regime that is similar to a typical regime
of supercapacitors operation.
Moreover, when impedance measurements are performed in
a non--linear regime
it is difficult\cite{kompan2010nonlinear} to interpret
measured $Z(\omega)$ into capacitance and internal resistance.
Since time domain interpretation is based on direct measurements of
charge preservation and energy dissipation
invariants --- it remains valid even in a highly non-linear regime.
The main disadvantage is that
in impedance spectroscopy the $\omega$
can capture a wide range (at lest seven orders) of frequencies,
but in time-domain measurement it is difficult
to capture more than three orders of $\tau$.

Consider Fig. \figureRef{cfigCircuit}{d} circuit.
Initially the switch ``Short'' is set to off, the switch ``Charge'' is set to on,
the supercapacitor is charging;
after a long enough time the switch ``Charge'' is set to off,
the supercapacitor is considered charged.
Then at $t=0$ the switch ``Short'' is set to on (shorting stage)
and the $I(t)$ current is measured in the external circuit.
At $t=\tau$ the switch ``Short'' is set to off, the current is interrupted.
The $R_s$ is a small resistance used to measure the current,
typically this is an internal resistance of wires and switches,
it can be determined\cite{kompan2021inverse}
using either four-terminal sensing technique
or calibrated to total charge.
Measuring $U_s$  we obtain the current
$U^*/R_s=I$. 
If there is only a single $RC$
then on shorting stage ($R_s\ll R$) we have:
\begin{align}
\frac{U^*}{R_s}=\frac{U}{R}=I&=C\frac{dU}{dt} \label{CurrentC}
\end{align}
For a single $RC$ circuit the values of $C$ and $R$ are exact values
not depending on shorting time $\tau$. For a multi-branch $RC$
circuits in Fig. \ref{cfigCircuit}
the values can be interpreted as some $\tau$-dependent
effective values $R(\tau)$ and $C(\tau)$ 
characterizing the supercapacitor. Integrating (\ref{CurrentC}),
we obtain:
\begin{align}
Q&=\int\limits_0^{\tau} I(t)dt=C\cdot(U_0-U_1) \label{QCurrentC}
\end{align}
The total charge passed is calculated
by integrating the current,
the limits of integration can be extended to be the entire timeline
since $I(t)=0$ for $t<0$ and $t>\tau$.
Initial and final potentials $U_0$ and $U_1$
can be measured as the potentials before shorting and right
after switching to open circuit regime.
Obtain
\begin{align}
C(\tau)&=\frac{Q}{U_0-U_1}
=\frac{\int I(t)dt}{U_0-U_1}
\label{CCurrentC} \\
R_1&=\frac{U_0}{I_0}=\frac{U_1}{I_1} \label{RiCurrentC}
\end{align}
The $C(\tau)$ can be interpreted as an effective
value characterizing the supercapacitor
at shorting time $\tau$.\footnote{
Thе $\tau$ is a probing parameter in time-domain spectroscopy,
not actual time $t$;
it is an analogue of probing frequency $\omega$ in
impedance spectroscopy. To compare the results
in time- and frequency- domains use (\ref{TauRelOmega}).
}
The potential jump
after shorting or after switching to the open circuit regime
gives the lowest possible internal resistance $R_1$
(when the $R_s$ is not small as $R_s\ll R_1$ then (\ref{RiCurrentC})
gives $R_s+R_1$).
The $U_0$ and $U_1$
are supercapacitor's potentials measured immediately before and immediately after shorting; the $I_0$ and $I_1$ are supercapacitor's external circuit
current measured immediately after and immediately before shorting.
The $\tau$-dependent effective capacitance (\ref{CCurrentC})
and $\tau$-independent minimal internal resistance (\ref{RiCurrentC})
are the result of \cite{kompan2021inverse}.

The effective $R(\tau)$
characterizes how deep the current pulse of $\tau$ duration penetrates
into the pores of supercapacitor material. Consider a single $RC$ circuit
\begin{align}
\begin{circuitikz}[line width=1pt,european resistors]
  \draw (0,2) node[anchor=east] {} to[short, -] (0,2)
  to[R=$R(\tau)$] (3,2) ;
  \draw (0,0) node[anchor=east] {} to[short, -] (3,0);
  \draw (0,0) to[C=$C(\tau)$] (0,2);
  \draw (1.5,0) to (1.5,-0.02) node[line width=0.4pt,ground]{};
  \draw (0,2) to node[anchor=east] {$U_0\rightarrow U_1$} ++(0,0) ;
  \draw (3,2) to[nos=$\mathrm{Short}$] (3,0) ;
\end{circuitikz}
\label{RCEstimation}
\end{align}
if the potential on $C$ decreases from $U_0$ to $U_1$
with a single
exponent evolution then  $U_1=U_0\exp \left(\frac{-\tau}{R(\tau)C(\tau)}\right)$
and the effective resistance is:
\begin{align}
\widetilde{R}(\tau)&=\frac{\tau}{C(\tau)\ln\frac{U_0}{U_1}}
\label{RCurrentR:Dasha}
\end{align}
Whereas the capacitance estimation (\ref{CCurrentC})
is exact as it is based on charge preservation,
the (\ref{RCurrentR:Dasha})
is just an estimation based on an assumption
of single exponent\footnote{
One can use a more advanced technique of 
 Lebesgue integral quadrature\cite{ArxivMalyshkinLebesgue}
 to estimate the distribution of relaxation rates.
} evolution of the potential;
this does not hold true for multi-branch supercapacitors.
We need $R(\tau)$ estimation that is based on an invariant.
For a regular capacitor and (\ref{RCEstimation}) circuit
the invariant is energy dissipation law $RI^2=\frac{1}{2}CdU^2/dt$,
it follows from the dynamic equation (\ref{CurrentC})
by multiplying it by $U$. Integrating, we obtain
\begin{align}
R(\tau)&=\frac{\left[U^2_0 - U^2_1\right]C(\tau)}
{2 \int I^2(t) dt}=
\frac{\left[U_0 + U_1\right] \int I(t) dt}
{2 \int I^2(t) dt}
\label{RCurrentR:Dasha:Invariant}
\end{align}
For a simple (\ref{RCEstimation}) circuit
with regular capacitor
it gives the exact value of $R$.
With multi-branch capacitor that undergo multi-exponent
dynamics the invariant no longer holds exactly. However it can be used
to obtain $R(\tau)$ --- an estimation of the effective internal resistance at $\tau$; it is determined from a condition of $\int I(t) dt$ and $\int I^2(t) dt$
relation.
From (\ref{RCurrentR:Dasha:Invariant}) we obtain
$R(\tau)$ asymptotes:
$R(\tau\to 0)=R_1$ and $R(\tau\to \infty)=const$.

Whereas the $R_1$ from (\ref{RiCurrentC}) characterizes the minimal possible internal resistance,
the (\ref{RCurrentR:Dasha:Invariant}) characterizes the resistance
of the whole system at $\tau$, it is an analogue
of impedance real part at $\omega=1/\tau$.
The idea is then to treat the $[R(\tau),\tau/C(\tau)]$
parametric plot
as ``real'' and ``imaginary'' parts of system impedance.

\begin{figure}

\includegraphics[height=7cm]{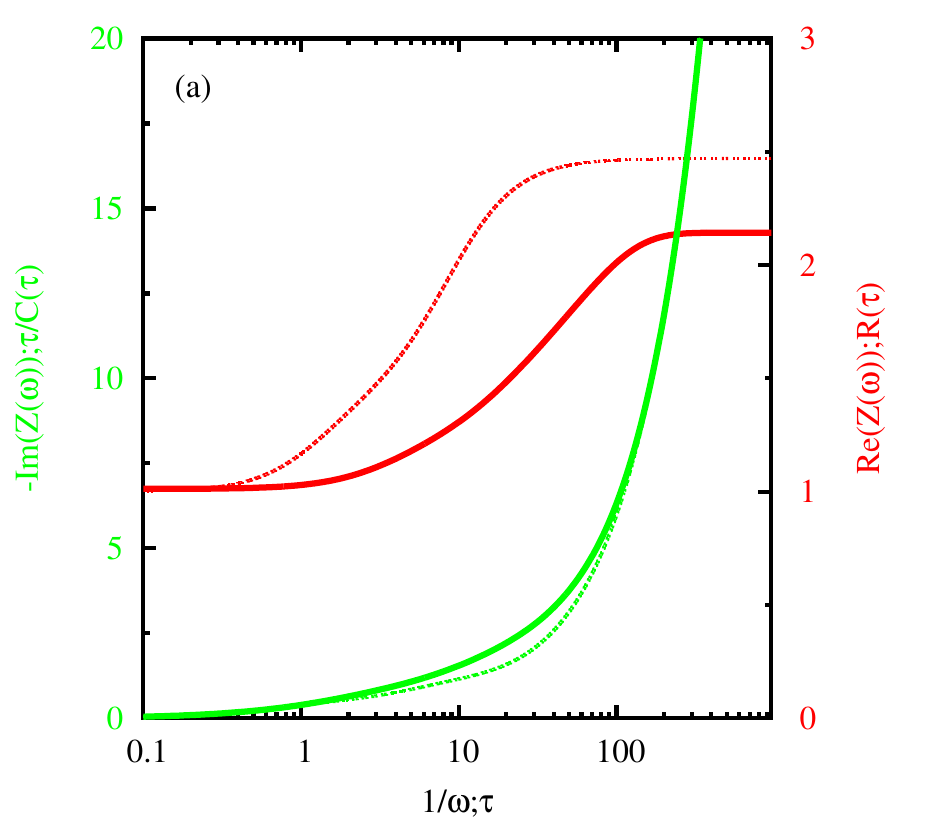}
\includegraphics[height=7cm]{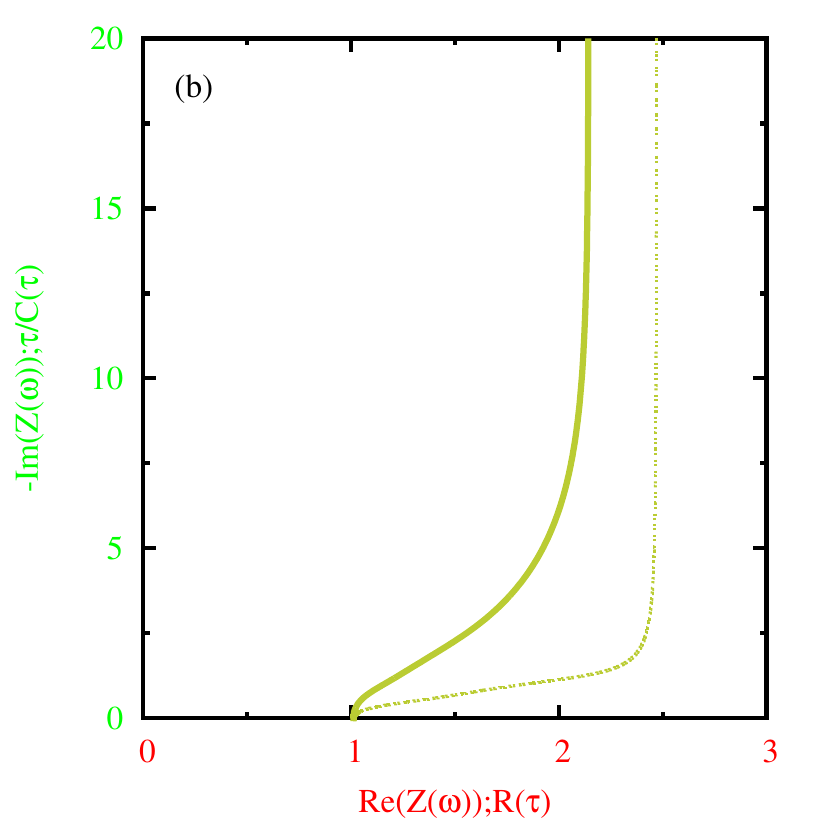}

\caption{\label{cfigCircuitModelling}
Modeling Fig. \figureRef{cfigCircuit}{a} circuit
with $R_1=1\Omega$, $C_1=2F$, $R_2=1\Omega$, $C_2=5F$, $R_3=2\Omega$, $C_3=10F$
in time (solid lines) and frequency (dashed lines, $\tau=1/\omega$) domains.
(a): real and imaginary part as a function on $\tau$; (b): parametric
Nyquist plot $[\mathrm{Re}(Z),\mathrm{Im}(Z)]$.
For time-domain the $R(\tau)$ and $\tau/C(\tau)$
are used instead of $\mathrm{Re}(Z)$ and $\mathrm{Im}(Z)$.
Note, that a divergence of $\mathrm{Im}(Z(\tau\to\infty))$
makes this representation less convenient than $R(\tau)$, $C(\tau)$
in Fig. \ref{cfigCircuitModellingCapacitance}.
  }
\end{figure}

In Fig. \ref{cfigCircuitModelling} we present the result
for a three-$RC$ system in Fig. \figureRef{cfigCircuit}{a}
modeled (see \ref{psiXSoftware} below)
in time (solid lines) and frequency (dashed lines, $\tau=1/\omega$) domains.
The imaginary part (green line) has the same $\tau\to 0$ and $\tau\to\infty$
asymptotes in both domains. A small difference is observed for
intermediate $\tau$. The real part (red line) has the same $\tau\to 0$
asymptote in both domains, but $\tau\to\infty$ asymptote is different.
The reason is that (\ref{RCurrentR:Dasha:Invariant})
averaging gives different (from impedance theory) weights to deep branches.

The basis $[\mathrm{Re}(Z),\mathrm{Im}(Z)]$ is not very convenient
for characterizing supercapacitors. First, the $\mathrm{Im}(Z)$ is not
bounded, it diverges
at $\tau\to\infty$ (corresponds to $\omega\to 0$)
if there is no self-discharge.
Second, accumulated energy
is proportional to $C$, thus it is convenient to present the $C$ explicitly.
For time--domain the $C(\tau)$ is given by (\ref{CCurrentC}),
for frequency domain let us define it as (\ref{CimpEq}).
There are other ways to introduce $C_{impedance}(\tau)$,
for example similar to Fig. \ref{differentRCcurcuits} above
we can introduce
$Y=1/Z$ and consider the value $\mathrm{Im}Y/\omega$.
However, this would be more \hyperref[parallelCircuit]{appropriate}
for the systems
with parallel $RC$; for supercapacitors the discharge is typically
small and the (\ref{CimpEq}) definition is reasonable. 
Thus the most convenient variables to characterize a supercapacitor from
measured impedance data $[\mathrm{Re}Z(\omega),\mathrm{Im}Z(\omega)]$
are\footnote{
\label{parallelCircuit}
These simple expressions (\ref{CimpEq}) and (\ref{RimpEq})
are applicable  only to the circuits
without self-discharge, where all $R$ are connected in serial.
For example for pure parallel connection \begin{circuitikz}[scale=0.3,transform shape,european resistors]
  \draw (0,1) node[anchor=east] {} to[short, *-] ++(0.5,0)
         to[short, -] ++(0,0.5) 
  to[R] ++(2.5,0) to [short, -] ++(0,-0.5) to[short, -*] ++(0.5,0);
  \draw (0.5,1)  to[short, -] ++(0,-0.5)
  to[C] ++(2.5,0) to [short, -] ++(0,+0.5) ;
\end{circuitikz}
the (\ref{CimpEq}) $C_{impedance}$ diverges at $\omega\to 0$;
correct answer is
$R=\mathrm{Re} Z\left[1+\left[\mathrm{Im} Z/\mathrm{Re} Z\right]^2\right]$,
$C=\frac{-1}{
\omega \mathrm{Im} Z \left[
1+\left[\mathrm{Re} Z/\mathrm{Im} Z\right]^2
\right] \label{CimpParEq}
}$.
Eq. (\ref{CimpEq}) matches this parallel circuit proper $C$ only
at high $\omega$.
There is a difficulty
in capacitance
estimation from impedance data with simple formula (\ref{CimpEq}), for proper results an equivalent circuit
and software modeling (e.g. in ZView) is required.
Time domain measurement technique (\ref{CCurrentC}) and (\ref{RCurrentR:Dasha:Invariant})
does not have a problem at
$\omega\to 0$ as it directly estimates capacitance
from measurement data.
}
\begin{align}
\tau &= \frac{1}{\omega} \label{TauRelOmega}\\
C_{impedance}(\tau)&=\frac{-1}{\omega\mathrm{Im}Z(\omega)}=
\frac{-\tau}{\mathrm{Im}Z(1/\tau)}
\label{CimpEq} \\
R_{impedance}(\tau)&=\mathrm{Re}Z(\omega)=\mathrm{Re}Z(1/\tau)
\label{RimpEq}
\end{align}
A similar approach to consider $[R,C]$ basis instead of
$[\mathrm{Re}Z(\omega),\mathrm{Im}Z(\omega)]$
was introduced in \cite{allagui2016spectral}
in application to supercapacitors
and in
\cite{ivanovNenashevAleshin2022low}
in application to perovskite-graphene
oxide composite films. Whereas for the films
the $[R,C]$ are just some characteristics of the material,
for supercapacitors the  $[R,C]$ are the most important
characteristics and presenting them
in the same plot 
gives a much better understanding of the device characteristics.

\begin{figure}
\includegraphics[height=4.7cm]{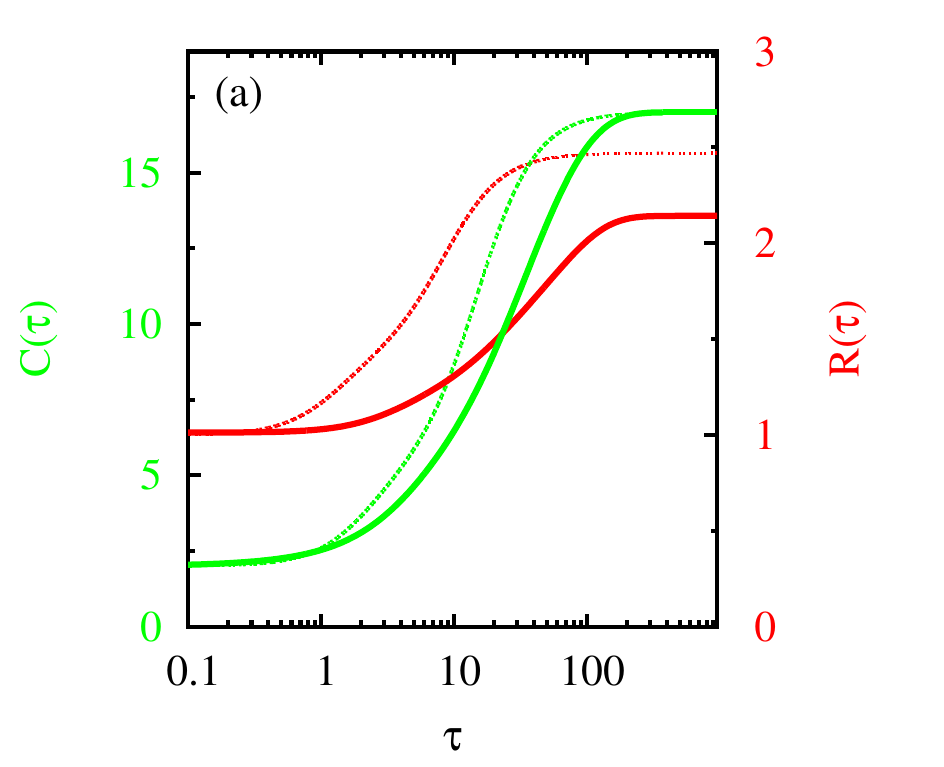}
\includegraphics[height=4.7cm]{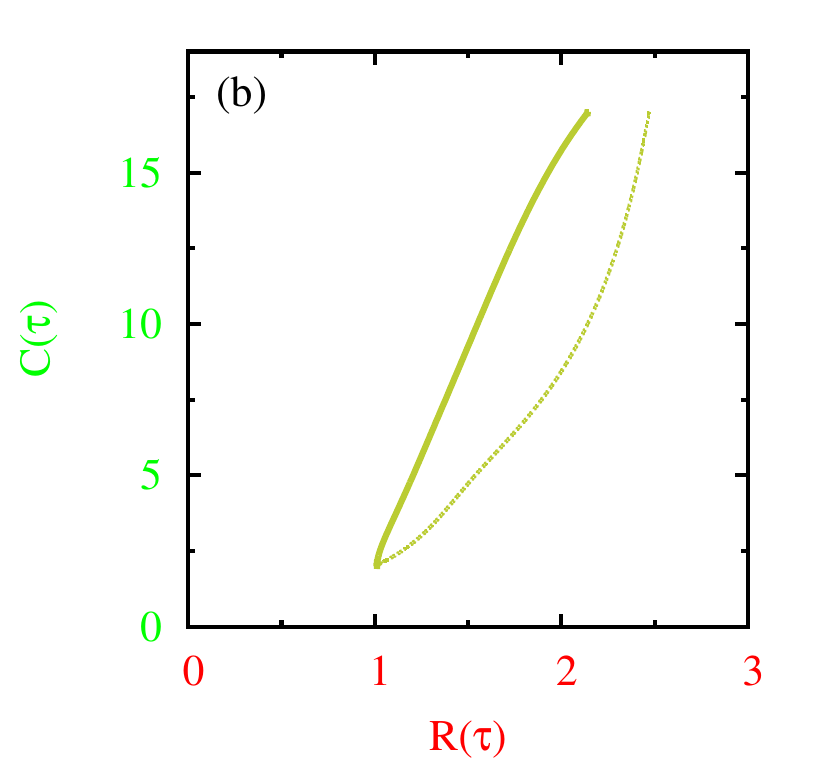}
\includegraphics[height=4.7cm]{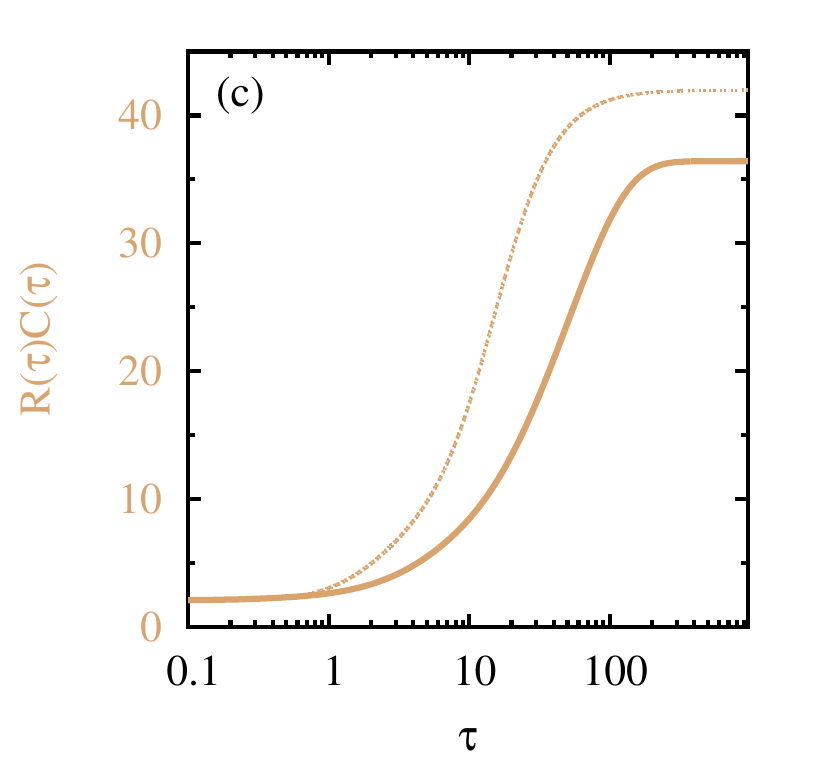}

\caption{\label{cfigCircuitModellingCapacitance}
Modeling Fig. \figureRef{cfigCircuit}{a} circuit
with $R_1=1\Omega$, $C_1=2F$, $R_2=1\Omega$, $C_2=5F$, $R_3=2\Omega$, $C_3=10F$
in time (solid lines) and frequency (dashed lines, $\tau=1/\omega$) domains.
(a): effective $R$ and $C$ as a function on $\tau$;
(b): parametric $[R(\tau),C(\tau)]$ plot that exhibits
close to linear behavior even in this simple three-$RC$ model;
(c): The effective internal $RC$ time as a function on $\tau$.
}
\end{figure}

In time-domain the values of $C(\tau)$ and $R(\tau)$
are obtained directly, no conversion necessary. The
expressions (\ref{CimpEq}) and (\ref{RimpEq})
of frequency domain correspond exactly to
(\ref{CCurrentC}) and (\ref{RCurrentR:Dasha:Invariant}) of time-domain.
In Fig. \ref{cfigCircuitModellingCapacitance} we present the result
for the same three-$RC$ system in Fig. \figureRef{cfigCircuit}{a}
 in $[R(\tau),C(\tau)]$ basis;
time domain is in solid line,
 frequency domain is in dashed line. As expected $\tau\to 0$ asymptotes
 match as $R(\tau\to 0)=R_1$  and $C(\tau\to 0)=C_1$; the $C(\tau\to \infty)$
 asymptotes
 matches exactly the total capacitance
 $C(\tau\to\infty)=C_1+C_2+C_3+\dots$,
 for $R(\tau\to\infty)$ the asymptotes are different in time and frequency domains.
 Most manufacturers provide equivalent ESR at fixed frequency $1000~\mathrm{Hz}$
in the datasheets,
which is typically several times lower than the internal resistance at DC.
In \cite{kompan2021inverse} a chart of supercapacitor's
internal $RC$ time
as a function of capacitance was made
for a number of supercapacitors
based on manufacturers datasheets.
In Fig. \figureRef{cfigCircuitModellingCapacitance}{c}
the $R(\tau)C(\tau)$ 
is presented as a function of shorting time $\tau$ for a single
supercapacitor, the dependence of supercapacitor
internal time on shorting
time $\tau$. If we divide accumulated energy $CU^2/2$ by current
power $U^2/R$ the ratio (within a factor of 2)
will be the supercapacitor internal time $R(\tau)C(\tau)$.
It characterizes the distribution of the internal porous structure of the system.

 The parametric plot $[R(\tau),C(\tau)]$
 (Fig. \figureRef{cfigCircuitModellingCapacitance}{b}, olive) is the most informative.
 It shows how the supercapacitor  behaves at different $\tau$.
 The $R(\tau)$ and $C(\tau)$ increase together with $\tau$ 
 as the
electric current penetrates into deeper and deeper pores of supercapacitor material.
This $[R(\tau),C(\tau)]$ parametric plot shows what energy and power can be possibly obtained from
a supercapacitor at a given time--scale $\tau$.
For device properties analysis
it is more convenient than regular
Nyquist plot
$[\mathrm{Re}Z,\mathrm{Im}Z]$ that requires equivalent circuit fitting.

\subsection{\label{modelingBranch}Time--Domain Modeling of Various Equivalent Circuits}

Equivalent circuits used to describe supercapacitors
electric properties typically contain
several building blocks shown in Figs. \figureRef{cfigCircuit}{a,b,c}.
It is well known \cite{bagotsky2015electrochemical,barsoukov2018impedance}
that different equivalent circuits can provide identical impedance behavior;
different SC models containing more than a dozen of $RC$ elements
create a good fit to almost any experimental data.

A transmission line model (horizontal ladder network)
 is very popular in the supercapacitors
 community\cite{fletcher2014universal,devillers2014review,logerais2015modeling,zhang2018review}.
It considers a number of  $RC$ elements connected in chain,
Fig. \figureRef{cfigCircuit}{a}.
Alternatively one can model a supercapacitor with a number of $RC$ elements
connected in parallel (superposition, vertical ladder network),
 Fig. \figureRef{cfigCircuit}{b}. Given sufficient number of $RC$ elements
both circuits fit various experimental data well, for example
Fig. \figureRef{cfigCircuit}{b} can be used to model
a \href{https://en.wikipedia.org/wiki/Constant_phase_element}{CPE}
element\cite{valsa2013rc}.
Good fitting with a complex circuit, however, does not guarantee that
a supercapacitor has exactly this equivalent circuit on material level.

A generalization of these models is multi-branch circuits
where an element at $k$-th level is connected to
more than one element at $k+1$-th level, a tree-like equivalent
circuit\cite{sen2018implicit,elwakil2021equivalent,elwakil2022generalizing}.
\href{https://en.wikipedia.org/wiki/Binary_tree}{Binary tree},
where $k$-th level node is connected to two
$k+1$-th level nodes (exactly two descendant nodes)
   is a simple example of multi--branch,
see Fig. \figureRef{cfigCircuit}{c},
it has exactly $2^k$ capacitors (and resistors) at
$k$-th level, totally $2^{k+1}-1$ capacitors (and resistors) on
all levels below or equal $k$. It is convenient
to numerate elements with two indexes: level $k$ and element number within
level $m=1\dots 2^k$.
If all the capacitors (and resistors) for a given level $k$
are the same, i.e. $C_{km}$ (and $R_{km}$) do not depend on $m$
then,
considering system symmetry, one can immediately obtain
an equivalence to transmission line model\cite{elwakil2021equivalent} with
\begin{subequations}
\label{crbinarytree}
\begin{align}
C_k&=C_{k*}\cdot 2^k \label{crbinarytreeC} \\
R_k&=R_{k*}/2^k \label{crbinarytreeR}
\end{align}
\end{subequations}
(for such a high symmetry system the potential
is the same for all capacitors in the $k$-th level).
A binary tree with $C_{km}=0$ and $R_{km}=0$ for all $k$
\textsl{except} the one equals to the tree depth (max level)
is equivalent to superposition model in Fig.  \figureRef{cfigCircuit}{b}.
For a binary tree with arbitrary $RC$,
as well as for a tree with varying number of descendant nodes,
this equivalence to Fig. \figureRef{cfigCircuit}{a,b}
simple models no longer holds.
Note, that
one can represent an arbitrary (e.g. having varying number of child nodes)
$RC$ network of
\href{https://en.wikipedia.org/wiki/Tree_(data_structure)}{tree structure}
with a binary tree by inserting dummy elements with properly chosen $R$ and $C$,
similarly to an equivalence of transmission line and superposition models
to a binary tree of special form.

\begin{figure}

\includegraphics[height=3.3cm]{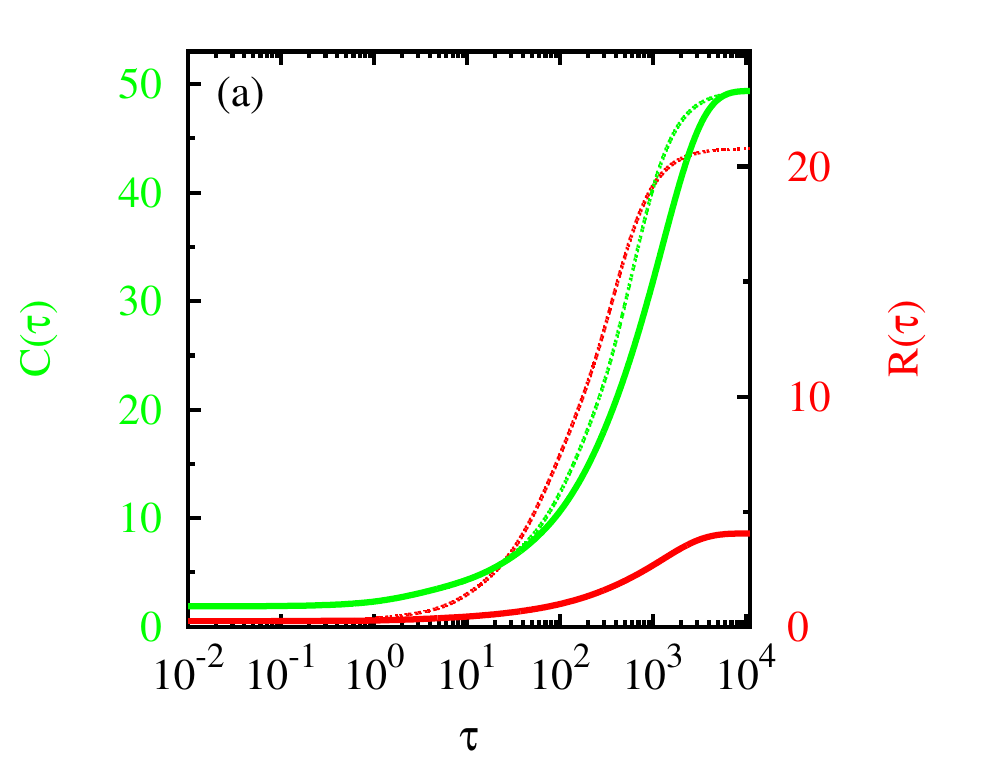}
\includegraphics[height=3.3cm]{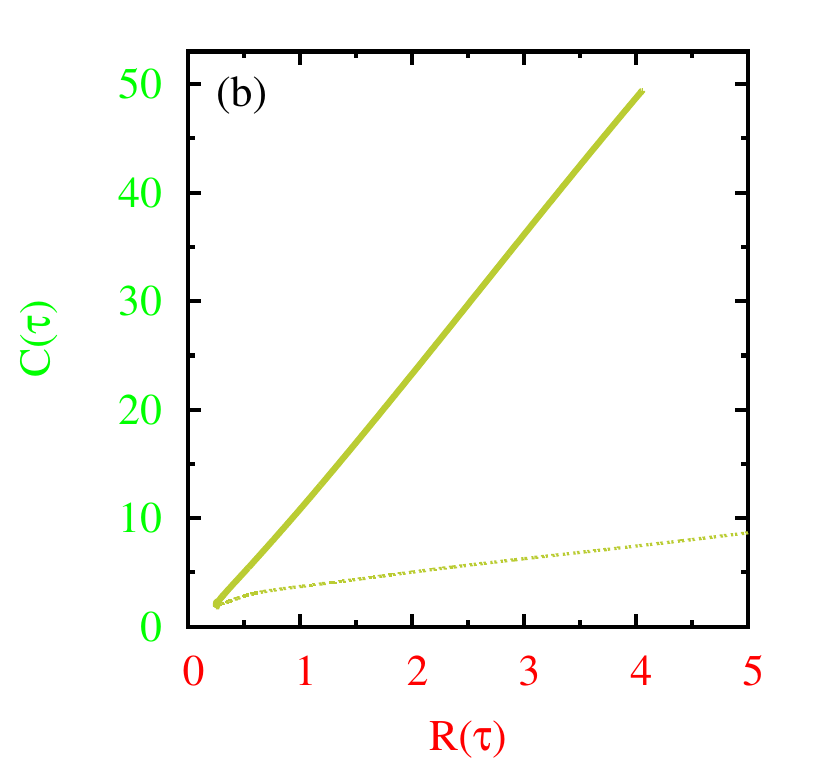}
\includegraphics[height=3.3cm]{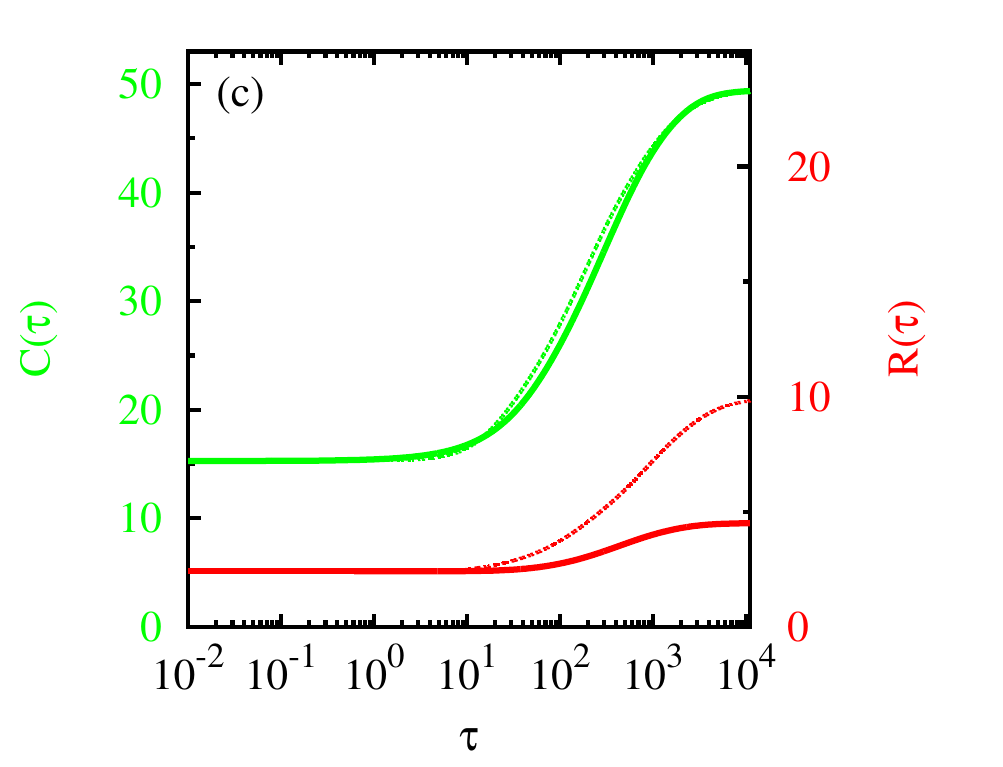}
\includegraphics[height=3.3cm]{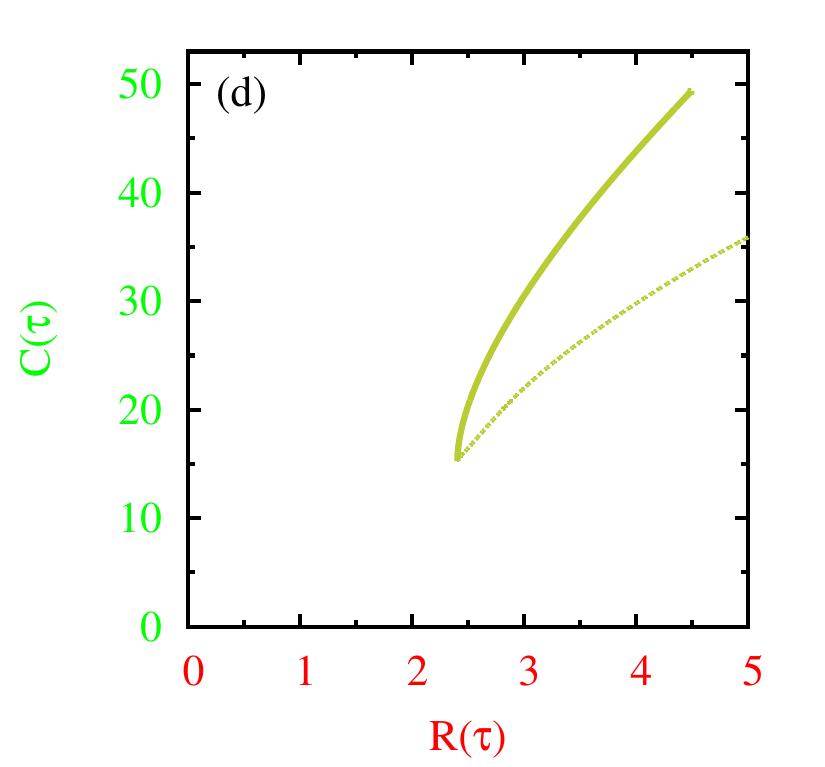} \\
\includegraphics[height=3.3cm]{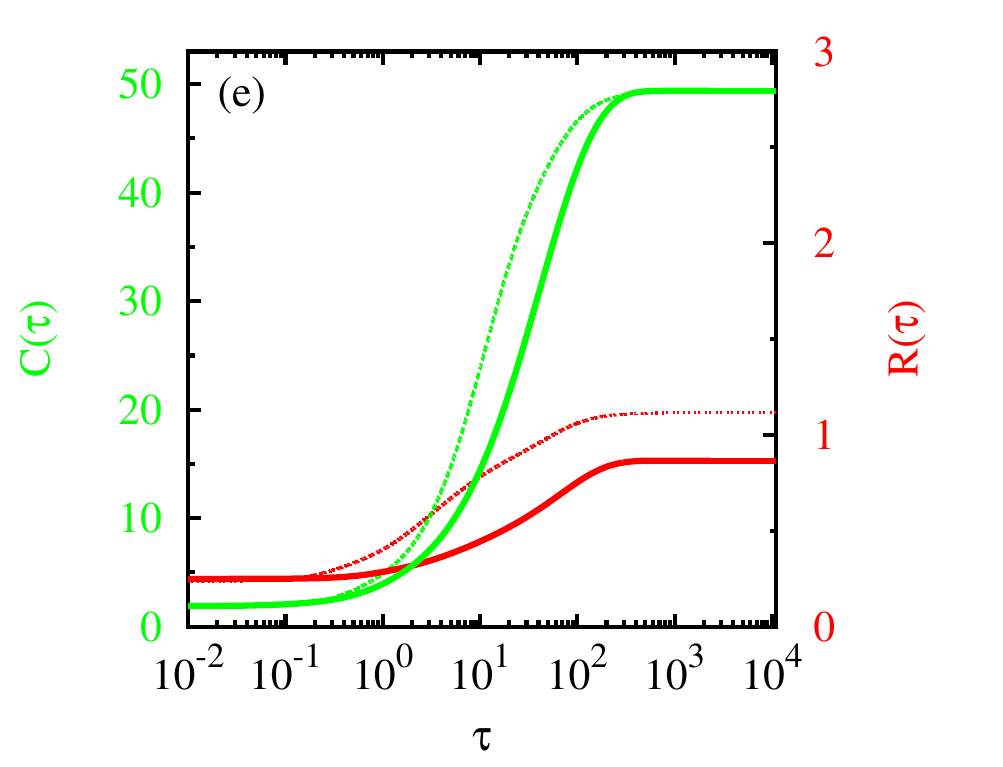}
\includegraphics[height=3.3cm]{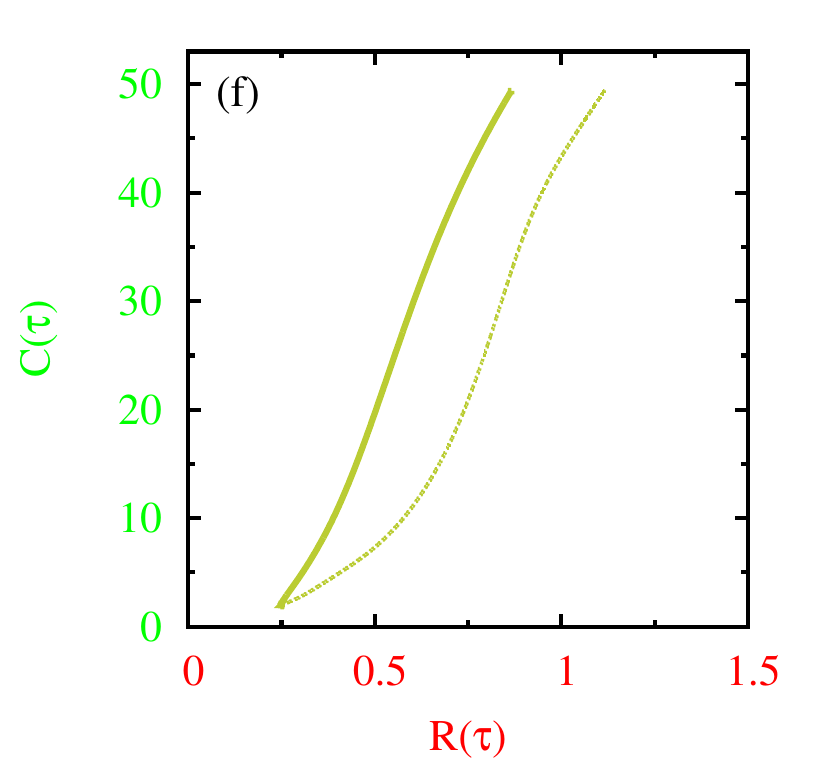}
\includegraphics[height=3.3cm]{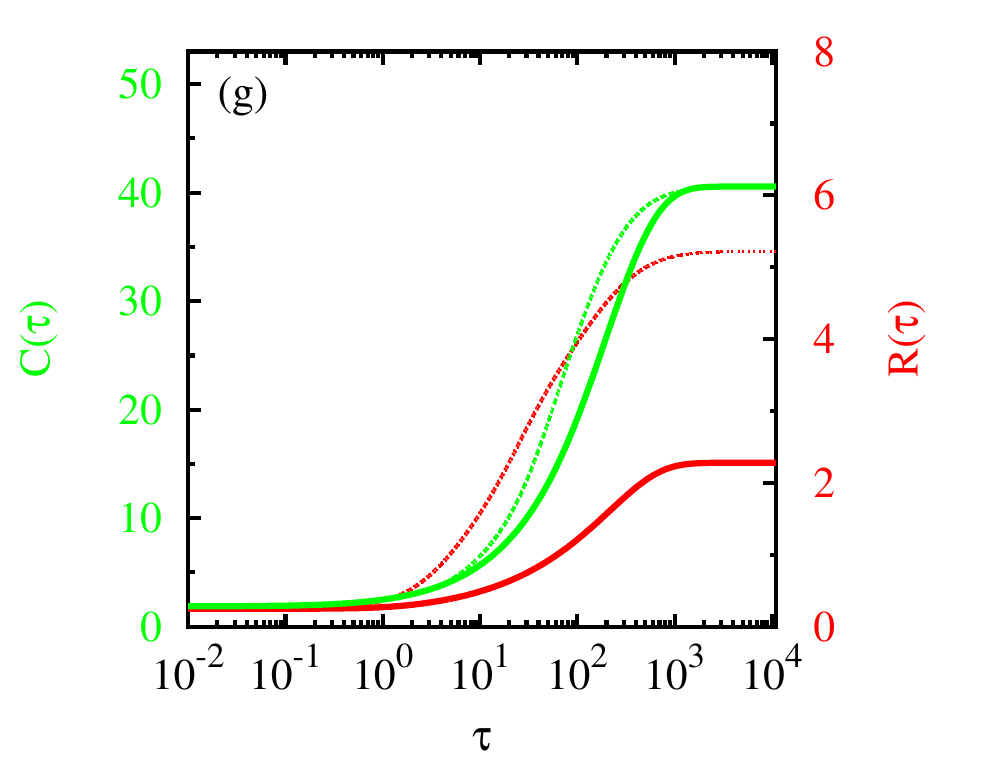}
\includegraphics[height=3.3cm]{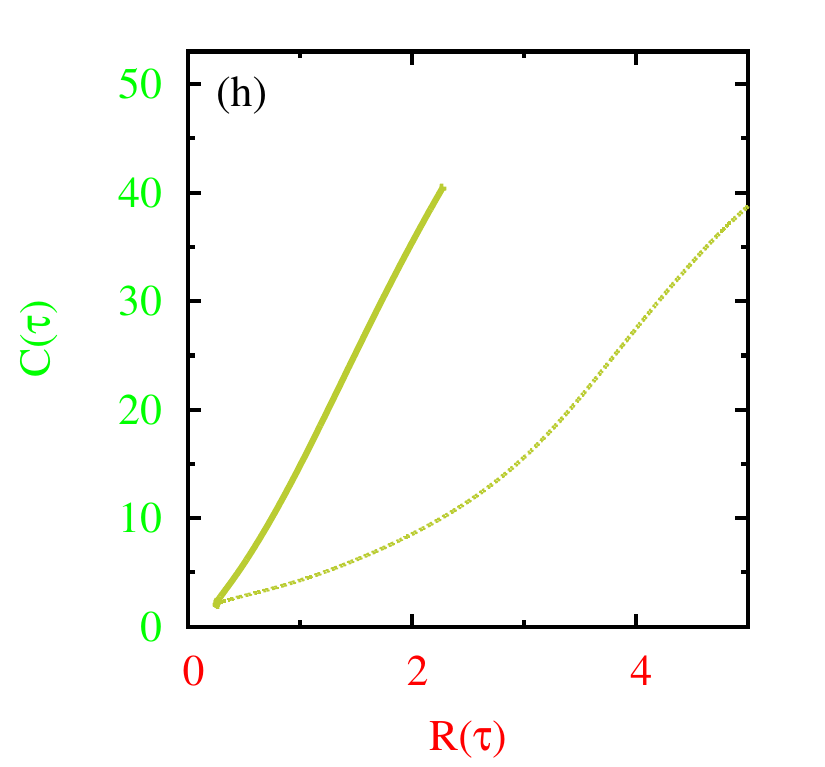} \\
\includegraphics[height=3.7cm]{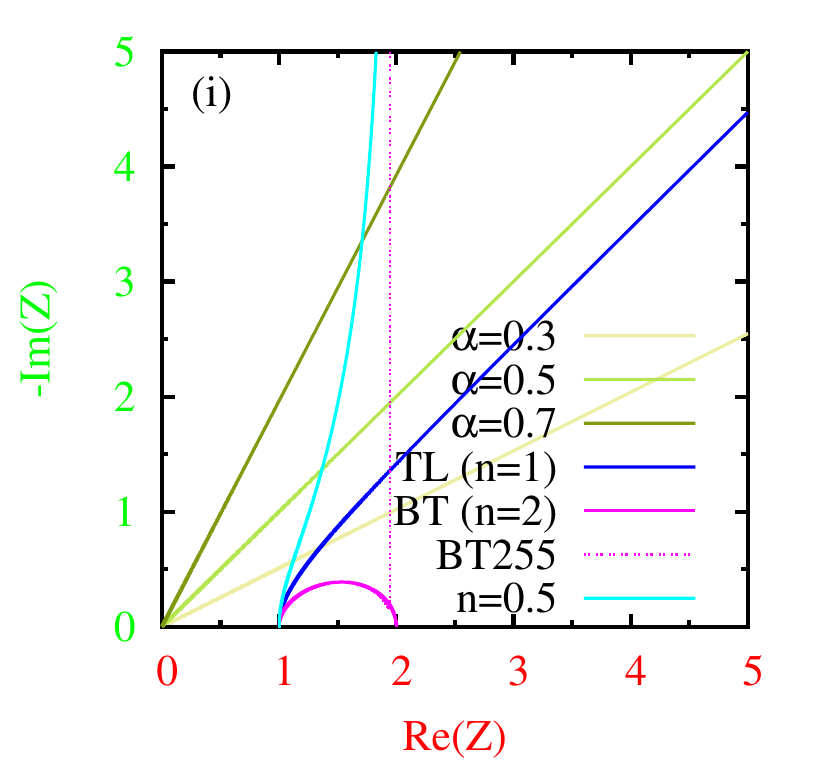}
\includegraphics[height=3.7cm]{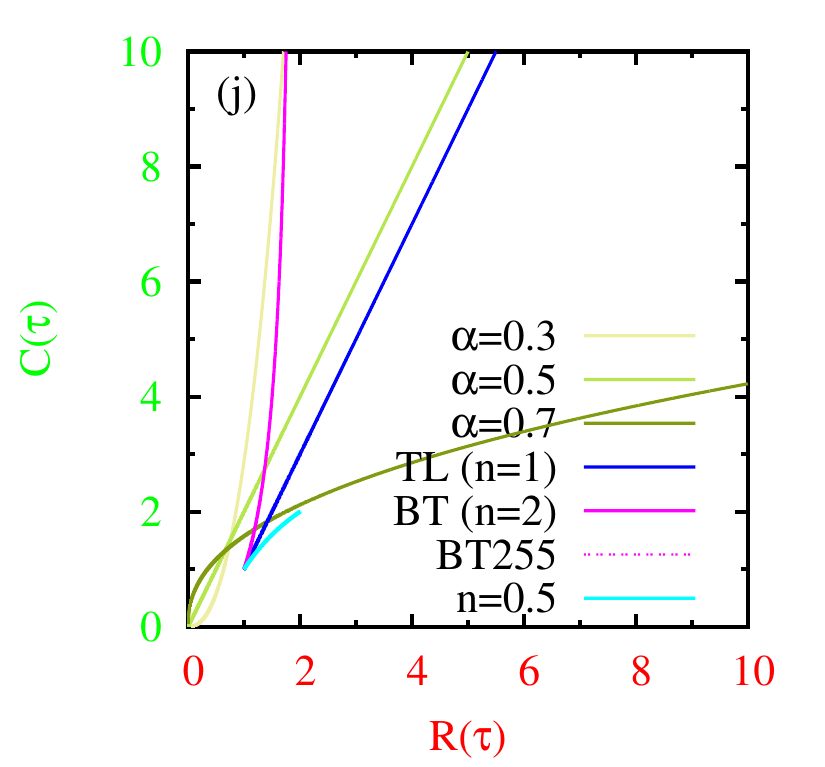}
\includegraphics[height=3.7cm]{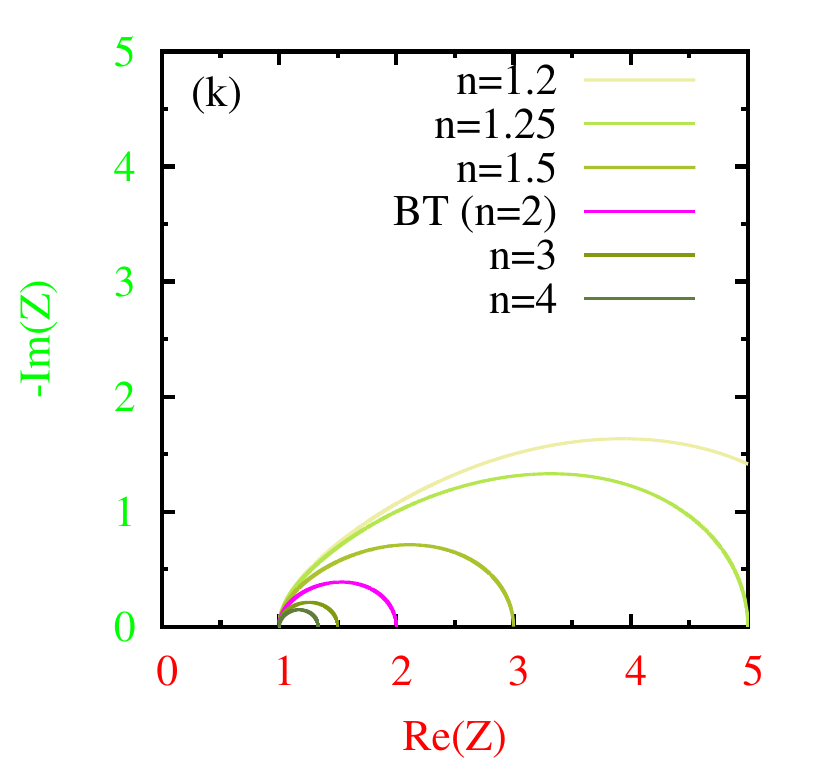}
\includegraphics[height=3.7cm]{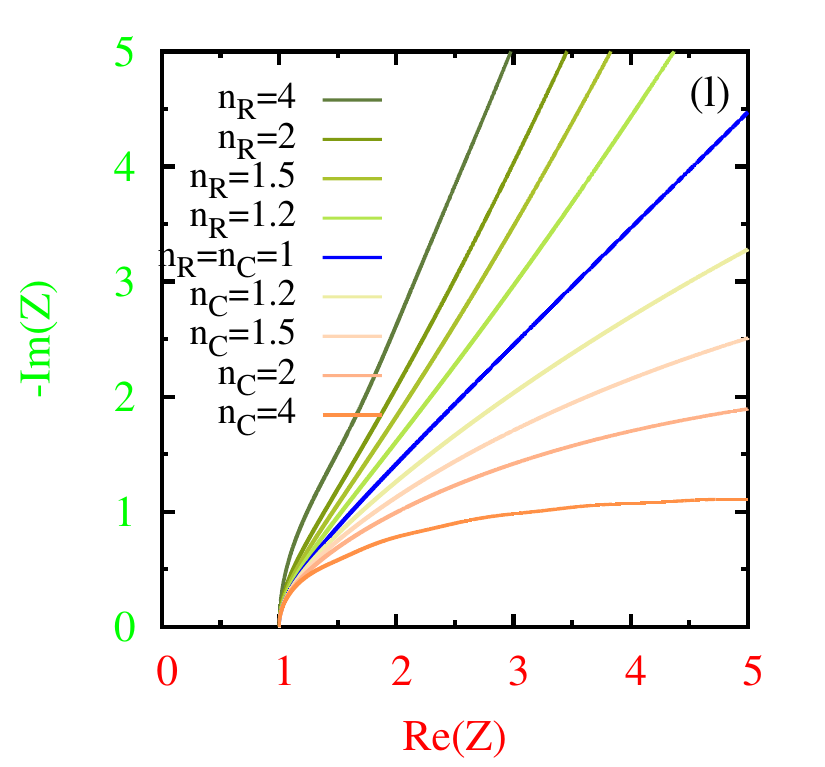}

\caption{\label{cfigCircuitModellingNetwork}
Modeling $RC$-network circuits. In time (solid line) and frequency (dashed line)
domains:
(a), (b) -- Fig. \figureRef{cfigCircuit}{a} transmission line model
with $31$ random $RC$ elements;
(c), (d) --  Fig. \figureRef{cfigCircuit}{b} superposition model
with $31$ random $RC$ elements;
(e), (f) --  Fig. \figureRef{cfigCircuit}{c} binary tree model of depth $4$, 
 $31$ random $RC$ elements;
(g), (h) --  Fig. \figureRef{cfigCircuit}{c} binary tree model of depth $7$, 
 $255$ random $RC$ elements.
Modeling in frequency domain to compare with CPE:
 (i), (j) -- CPE element with $\alpha=\{0.3,0.5,0.7\}$ (\ref{CPEZw}),
 infinite transmission line \textbf{TL} (\ref{ZChainValue}),
  binary tree --- infinite \textbf{BT} (\ref{ZChainValueBT}), of $255$ identical $RC$ elements \textbf{BT255},
  and $n=0.5$ weakly linked n-tree (\ref{Zsol}).
(k) nTE -- n-tree element (\ref{Zsol})
exhibits deformed semi-circles of Cole-Cole style
for $n>1$.
(l) CPE modeling with self-similar $RC$ network (\ref{RCdep})
of $100$ $RC$ elements
for different $n_C$, $n_R=1$ when $\alpha<0.5$,
and $n_R$, $n_C=1$ when $\alpha>0.5$, see \ref{AppndDifferentRCexponent}.
  }
\end{figure}

To create $RC$ networks in Fig. \ref{cfigCircuit}
we need to define a distribution for $R$ and $C$.
In this paper the
\href{https://en.wikipedia.org/wiki/Log-normal_distribution}{log-normal}
distribution is used for the reason to minimize the number of parameters
and to avoid the problem
with negative values when normal distribution is used.
For a number of $RC$ networks we plot $R(\tau)$, $C(\tau)$,
and $[R(\tau),C(\tau)]$ parametric plot
in time and frequency domains. The
$[\mathrm{Re} Z(\omega),\mathrm{Im} Z(\omega)]$
plots are much less informative since $\mathrm{Im} Z(\omega)$
diverges at $\tau\to\infty$, see Fig. \ref{cfigCircuitModelling}.
The $C(\tau)$ has known $\tau\to0$ and $\tau\to \infty$ asymptotes.
If there is no discharge --
they are identical
both in time and frequency domains
 as well as the $R(\tau\to 0)$.
The $C(\tau\to\infty)$ is the sum of all $C$ of the network.

In Fig. \figureRef{cfigCircuitModellingNetwork}{a,b}
the transmission line model, Fig. \figureRef{cfigCircuit}{a},
of $31$ random $RC$ elements is presented;
the value of $31$ is chosen to simplify the comparison below with
binary tree model of depth $k=4$ that has
$2^{k+1}-1$ capacitors (and resistors).
The network is built as
$C=\exp(\mathcal{N}(0,1))$, $R=\exp(\mathcal{N}(0,1))$, 
where $\mathcal{N}(0,1)$ is Gaussian random variable
with zero mean and unit variance, standard normal distribution.
We observe almost perfect linear dependence in $[R(\tau),C(\tau)]$ parametric plot.
The $C/R$ slope is different in time and frequency domains.
Linear dependence is observed by six orders of $\tau$,
but only four orders of $\tau$ range are informative as a plateau in
$R(\tau)$ and $C(\tau)$ is reached outside of the range.

This  $[R(\tau),C(\tau)]$ parametric plot\footnote{
As we consider a parametric plot parameter transform does not change
the plot, parametric plots
$[R(\tau),C(\tau)]$ and $[R(\omega),C(\omega)]$
are identical.
}
linear behavior
is a general property of transmission line model.
An analytic solution in frequency domain can be obtained in some cases.
Consider Fig. \figureRef{cfigCircuit}{a} model
of infinite length of identical $RC$: $R_k=R$ and $C_k=C$.
Since adding one more $RC$ element to an infinite chain
does not change the $Z(\omega)$, we obtain:
\begin{align}
Z&=R +\frac{Z\frac{1}{j\omega C}}
{Z+\frac{1}{j\omega C}} \label{Zeq} \\
Z(\omega)&=\frac{R}{2}\pm\sqrt{\frac{R^2}{4}+\frac{R}{j\omega C}}
\label{ZChainValue} \\
\frac{\Delta C}{\Delta R} &= \frac{\frac{dC_{impedance}}{d\omega}}{\frac{dR_{impedance}}{d\omega}} =
\frac{\displaystyle\frac{d}{d\omega} \frac{-1}{\omega \mathrm{Im}Z(\omega)}}
{\displaystyle\frac{d}{d\omega} \mathrm{Re}Z(\omega)} \label{CRslopeW} \\
\frac{\Delta C}{\Delta R} &=\frac{2C}{R} \label{CRw}
\end{align}
Impedance $Z(\omega)$ of Fig. \figureRef{cfigCircuit}{a}
infinite chain of identical $RC$
is obtained as quadratic equation
solution;
the solution corresponds to ``$+$'' sign in  (\ref{ZChainValue})
when the square root operation is defined as having positive real part.
From (\ref{ZChainValue}),
after simple algebra and symbolic calculations,
see \ref{SymCalc} below,
it immediately follows that
in variables $R_{impedance}(\tau)$ from (\ref{RimpEq})
and $C_{impedance}(\tau)$ from (\ref{CimpEq})
the parametric plot $[R_{impedance}(\tau),C_{impedance}(\tau)]$
for transmission line of infinite length
is linear
with $C/R$ slope equals \textsl{exactly} to $2C/R$ from (\ref{CRw}):
\begin{align}
C_{impedance}(\tau)&=\frac{2C}{R}R_{impedance}(\tau)-C
\label{TLinfLength}
\end{align}
A close to linear law also holds for a finite length
transmission line with $RC$ elements having
randomness, see Fig. \figureRef{cfigCircuitModellingNetwork}{b}.
When a finite length transmission line is considered in 
variables $[\mathrm{Re} Z(\omega),\mathrm{Im} Z(\omega)]$,
as it is typically done for SC,
the $\mathrm{Im}Z(\omega)$ diverges
at $\omega\to 0$
as in Fig. \figureRef{cfigCircuitModelling}{b}
and the system is hard to identify.
This non-conformal mapping
from $[\mathrm{Re} Z(\omega),\mathrm{Im} Z(\omega)]$
to $[R(\tau),C(\tau)]$
allows us easily identify a transmission line:
if there is a linear dependence in $[R(\tau),C(\tau)]$
parametric plot -- there is a transmission line model
corresponding to this $Z(\omega)$.
For time-domain consideration we cannot obtain an analytic
solution. Numerical experiments with very long transmission lines
(from $100$ to $500$ identical $RC$ elements)
show linear dependence in time domain, the
$[R(\tau),C(\tau)]$ parametric plot
is almost linear by five orders of $\tau$ range,
 Eqs. (\ref{CCurrentC}) and (\ref{RCurrentR:Dasha:Invariant}).
The $C/R$ slope is higher than the one in frequency domain,
in time domain
it varies from $4.8C/R$ at low $\tau$ to $7.5C/R$ at high $\tau$,
compare this with the exact value of $2C/R$ in frequency domain (\ref{CRw}).
This leads us to conclude that linear $[R(\tau),C(\tau)]$
parametric plot is an intrinsic property of a transmission line model.

Exactly linear $C/R$ slope is not unique to transmission line
with $Z(\omega)$ from (\ref{ZChainValue}).
Consider \href{https://en.wikipedia.org/wiki/Constant_phase_element}{CPE}
element\cite{bisquert2002theory,valsa2013rc} in frequency domain. It has
\begin{align}
Z(\omega)&=\frac{1}{ (j\omega)^{\alpha} C_\alpha}
\label{CPEZw}\\
-1&\le\alpha\le 1 \label{alphaC}
\end{align}
The case $\alpha=1$ corresponds to regular capacitance,
$\alpha=-1$ corresponds to inductance.
In $[\mathrm{Re}(Z),\mathrm{Im}(Z)]$ parametric plot we have a line
with the slope determined by the value of $\alpha$, see Fig. \figureRef{cfigCircuitModellingNetwork}{i}.
Let us apply non-conformal mapping to obtain 
$R_{impedance}(\tau)$ from (\ref{RimpEq}) and $C_{impedance}(\tau)$ from (\ref{CimpEq})
for different values of $\alpha$. The result is presented in Fig. \figureRef{cfigCircuitModellingNetwork}{i,j}. One can clearly see that $\alpha=0.5$
CPE case (diffusion limited process) has exactly linear dependence
in $[R(\tau),C(\tau)]$ parametric plot. The slope is the same as
for transmission line Eq. (\ref{ZChainValue}):
compare blue $n=1$ line (infinite transmission line with $R=1\Omega$, $C=1F$)
and olive line of $\alpha=0.5$ CPE element with $C_{\alpha}=1$ .
Transmission line is a good model\cite{bisquert2002theory} for diffusion-limited
CPE with $\alpha=0.5$;
a shift is due to $R/2$ term in (\ref{ZChainValue}).

In Fig. \figureRef{cfigCircuitModellingNetwork}{c,d}
the superposition model, Fig. \figureRef{cfigCircuit}{b},
of $31$ random $RC$ elements is presented. The network is built as
$C=\exp(\mathcal{N}(0,1))$, $R=100\exp(\mathcal{N}(0,1))$,
a factor of $100$ is chosen to bring the effective $R$ of the network
to approximately the same range as for other models,
note that $R(\tau\to 0)=1\big/\sum\limits_{k} 1/R_k$.
This Fig. \figureRef{cfigCircuit}{b}
model has the most noticeable deviation from linearity in
$[R(\tau),C(\tau)]$ parametric plot since 
it has no deep branches.
Linear dependence is observed by a singe order of $\tau$ range.
At small $\tau$ the $C(R)$ plot convexity (deviation from linear law)
for the model
has an \textsl{opposite sign}
than for the experimental data
in Fig. \ref{compareTW}.
For a superposition model with identical elements
an analytic solution can be obtained
both in time and frequency domains. The system with $n$ identical
$\widetilde{R}\widetilde{C}$
elements 
is equivalent to a single $RC$ of (\ref{RCEstimation}) form
with $R=\widetilde{R}/n$ and $C=\widetilde{C}n$.
This single $RC$ circuit has no distributed branches and 
its parametric plot $[R(\tau),C(\tau)]$ is a \textsl{single}
point $[R,C]$ both in time and frequency domains.

In Fig. \figureRef{cfigCircuitModellingNetwork}{e,f}
a binary tree model, Fig. \figureRef{cfigCircuit}{c},
of depth four ($31$ random $RC$ elements) is presented. The network is built as
$C=\exp(\mathcal{N}(0,1))$, $R=\exp(\mathcal{N}(0,1))$.
The
model has little deviation from linearity and, contrary
to transmission line results in Fig. \figureRef{cfigCircuitModellingNetwork}{a,b},
the $C/R$ slope is very similar in time and frequency domains;
similar behavior has also been observed in experimental data in Fig. \ref{compareTW}. At small $\tau$ the $C(R)$ plot convexity for the model
has  the \textsl{same sign}
as in the experimental data.

In Fig. \figureRef{cfigCircuitModellingNetwork}{g,h}
the binary tree model, Fig. \figureRef{cfigCircuit}{c},
of depth seven ($255$ random $RC$ elements) is presented;
this value is chosen to be able to consider depth-dependent
factors while in the same time not to have too many $RC$ elements
to avoid any possible numerical instabilities.
The network is built as:
\begin{subequations}
\label{betaFact}
\begin{align}
C_{km}&=\exp(\mathcal{N}(0,1))/\beta_C^k \label{cBetaC}\\
R_{km}&= \exp(\mathcal{N}(0,1))\beta_R^k \label{cBetaR}
\end{align}
\end{subequations}
Different for $C$ and $R$ depth-dependent factors $\beta_C$ and $\beta_R$
are introduced to construct a more realistic $RC$ distribution.
They partially compensate growing number of elements
with tree depth increase.
If $\beta_C=\beta_R=2$ then these factors totally compensate
(\ref{crbinarytree}) and this $RC$ network becomes similar to the
 transmission line model.
In Fig. \figureRef{cfigCircuitModellingNetwork}{g,h}
we put $\beta_C=1.5$ and $\beta_R=1.8$, thus only partial compensation takes place.
The $C/R$ is still almost linear in time-domain and there is a noticeable
difference from $C/R$ value in frequency domain, a behavior
we already observed in random transmission line model
in Fig. \figureRef{cfigCircuitModellingNetwork}{a,b}.

For an infinite binary tree with identical $RC$ elements an analytic
solution in frequency domain can be obtained.
Since adding one more $RC$ level to an infinite binary tree
does not change the $Z(\omega)$, we obtain a recurrent relation
that leads to quadratic equation solution:
\begin{align}
Z&=R +\frac{Z\frac{1}{2j\omega C}}
{Z/2+\frac{1}{j\omega C}} \label{ZeqBT} \\
Z(\omega)&=
\frac{1}{2}\left[R-\frac{1}{j\omega C}\right]
\pm
\sqrt{\frac{1}{4}\left[R-\frac{1}{j\omega C}\right]^2+
\frac{2R}{j\omega C}}
\label{ZChainValueBT}
\end{align}
The calculation of $C/R$ slope Eq. (\ref{CRslopeW}),
however, does not give a constant as
it is for transmission line model in Eq. (\ref{CRw}),
the $C/R$ slope changes slightly with $\omega$.
Overall the result is similar 
to a binary tree with log-normal distribution of $R$ and $C$.
The exact result is 
shown in Fig. \figureRef{cfigCircuitModellingNetwork}{i,j} in pink:
for an infinite binary tree \textbf{BT} (solid), for a binary tree
of depth $7$ ($255$ identical $RC$)
\textbf{BT255} (dashed); they differ only at very small $\omega$.
A remarkable feature of the binary three model is that a Cole-Cole style
semi-circle can be obtained without charge leak, see
Fig.  \figureRef{cfigCircuitModellingNetwork}{i}
and compare it with
Fig. \figureRef{differentRCcurcuits}{b};
this binary tree corresponds to the transmission line
with $RC$ values depending on $k$ as Eq. (\ref{crbinarytree}).
The  $[\mathrm{Re} Z(\omega),\mathrm{Im} Z(\omega)]$
plot
is a deformed semi-circle with
$\mathrm{Re} Z(\omega\to\infty)=R$,
$\mathrm{Re} Z(\omega\to 0)=2R$,
and
$\mathrm{Im} Z(\omega\to\infty)=\mathrm{Im} Z(\omega\to 0)=0$
asymptotes. The $\mathrm{Im} Z(\omega\to 0)=0$ holds only for
an infinite tree, for a finite size tree it starts showing 
a capacitance-like behavior at some low $\omega$,
see  \textbf{BT255} dashed pink line in Fig. \figureRef{cfigCircuitModellingNetwork}{i}.

One can generalize Eqs. (\ref{Zeq}) and (\ref{ZeqBT}) recurrence
to a tree with an arbitrary number of descendants $n$:
\begin{align}
Z&=R +\frac{Z\frac{1}{nj\omega C}}
{Z/n+\frac{1}{j\omega C}} \label{ZeqBTn} \\
Z(\omega)&=
\frac{1}{2}\left[R+\frac{1-n}{j\omega C}\right]
\pm \sqrt{
\frac{1}{4}\left[R+\frac{1-n}{j\omega C}\right]^2
+\frac{Rn}{j\omega C}
} \label{Zsol}
\end{align}
We name this infinite $RC$-network of identical $RC$ elements
as the \textbf{n-tree element} (nTE),
this is a special case of a general tree-like system\cite{sen2018implicit}.
The $n=1$ corresponds to transmission line model (single descendant),
$n=2$ corresponds to binary tree model (two descendants).
The value of $n$ determines tree growth exponent.
In quadratic
equation solution the sign
is ``$+$'' in  (\ref{Zsol})
when the square root operation is defined as having positive real part.
A weakly linked ($n<1$) n-tree with $n=0.5$ is presented in
Fig. \figureRef{cfigCircuitModellingNetwork}{i,j} in light blue;
this model has bounded total $C$, thus
$C(\tau\to\infty)$ 
does not diverge for an infinite $RC$ network.
nTE asymptotes for $n<1$ are:
$R_{impedance}(\tau\to \infty)=R/(1-n)$,
$C_{impedance}(\tau\to \infty)=C/(1-n)$,
$R_{impedance}(\tau\to 0)=R$,
$C_{impedance}(\tau\to 0)=C$.
The $[R_{impedance}(\tau),C_{impedance}(\tau)]$ parametric plot
is close to linear
 but not exactly. Average $C/R$ slope for $n<1$
 is approximately equal to $C/R$
 what is about twice lower than the value in $n=1$ case, Eq. (\ref{CRw}).
At $n=1$ tree-like $RC$ network in question has a
percolation
\href{https://en.wikipedia.org/wiki/Phase_transition}{phase transition}
observed in $C(\tau\to\infty)$
divergence; at $n\ge 1$ the total capacitance of nTE becomes infinite,
i.e. the capacitance becomes limited by the device actual size.
Percolation properties of supercapacitor electrodes are actively studied
in recent works
 \cite{king2012percolation,vasilyev2019connections,lei2021n,goodwin2022gelation}.

For $n> 1$ the n-tree model
produces deformed semi-circle
in $[\mathrm{Re} Z(\omega),\mathrm{Im} Z(\omega)]$
with asymptotes:
$\frac{\Delta C}{\Delta R}\Big|_{\omega\to\infty \mkern-36mu}=(n+1)\frac{C}{R}$,
$\mathrm{Re} Z(\omega\to\infty)=R$,
$\mathrm{Re} Z(\omega\to 0)=\frac{n}{n-1}R$,
and
$\mathrm{Im} Z(\omega\to\infty)=\mathrm{Im} Z(\omega\to 0)=0$;
the $\mathrm{Im} Z(\omega\to 0)=0$ holds only
for an infinite $RC$ network.
In Fig. \figureRef{cfigCircuitModellingNetwork}{k}
we present n-tree  models
with different values of $n>1$. Deformed semi-circles
are clearly observed.
Since
carbon structures of supercapacitor electrodes
are hierarchical tree-like structures
this leads us to conclude that
 SC impedance semi-circles as in
Fig. \figureRef{differentRCcurcuits}{b}
on material level
can be explained by a n-tree model with $n>1$.
The n-tree model of identical $RC$ elements
is equivalent to a transmission line model
in Fig. \figureRef{cfigCircuit}{a}
with $k$-dependent\footnote{
When $n<1$ one can easily obtain total capacitance
$C(\tau\to \infty)=C/(1-n)$
as \href{https://en.wikipedia.org/wiki/Geometric_progression}{geometric progression} sum;
this holds true both in time and frequency domains.
A transmission line with $C_k$ and $R_k$
as two geometric progressions with different common ratio
can be used to model \cite{sugi2002frequency} a given $\alpha$ CPE,
see \ref{AppndDifferentRCexponent} below.
For an example of 3D self-similar $RC$ network see \cite{arbuzov2009three}.
} $C_k=n^kC$ and $R_k=R/n^k$.
See \ref{SymCalc} below for symbolic and numerical calculation of
$Z(\omega)$ from (\ref{Zsol}), the command
``\texttt{\seqsplit{python3\ n-tree\_element.py\ 1.25}}''
calculates n-tree impedance for a given number of descendants $n=1.25$.

The modeling above leads us to conclude
that linear $C/R$ behavior can be observed in various $RC$ networks
with deep branches. The systems without deep branches,
such as in Fig. \figureRef{cfigCircuit}{b},
have noticeable deviation from the linear law.
The C/R slope in time- and frequency- domains
is similar in some systems (such as binary tree) and
significantly different in others (such as transmission line).
Time domain technique uses (\ref{CCurrentC}) and (\ref{RCurrentR:Dasha:Invariant})
to directly measure the $C(\tau)$ and $R(\tau)$;
frequency domain technique measures the impedance $Z(\omega)$ first,
then uses Eqs. (\ref{CimpEq}) and (\ref{RimpEq})
to convert impedance data to effective $C$ and $R$;
the limitations of Eqs. (\ref{CimpEq}) and (\ref{RimpEq})
 makes the result much less accurate.
However, for some systems, e.g. binary tree modeling
in  Fig. \figureRef{cfigCircuitModellingNetwork}{e,f}
and experimental data in Fig. \ref{compareTW},
the $C/R$ slope is very similar in time and frequency domains.
A minimal system to observe a close to linear 
$[R(\tau),C(\tau)]$ dependence
is Fig. \figureRef{cfigCircuit}{a} transmission line model
with three $RC$ elements, see Fig. \figureRef{cfigCircuitModellingCapacitance}{b}.

\section{\label{timeDomainExperiments}Supercapacitors Experimental Measurements in Time Domain}

\begin{figure}

\includegraphics[height=4.7cm]{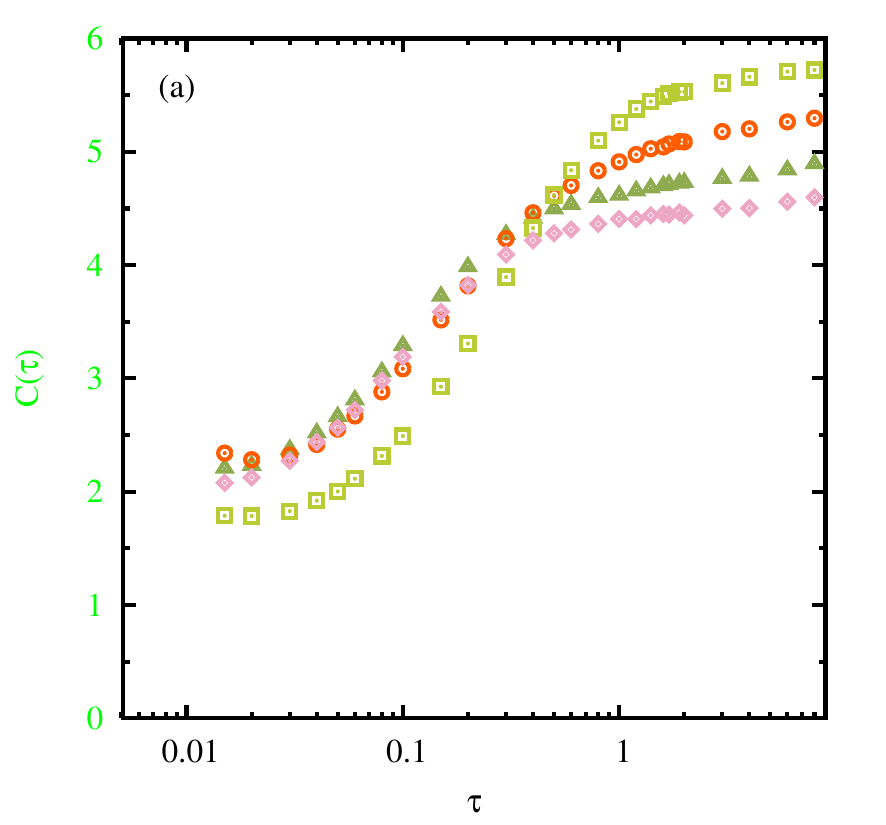}
\includegraphics[height=4.7cm]{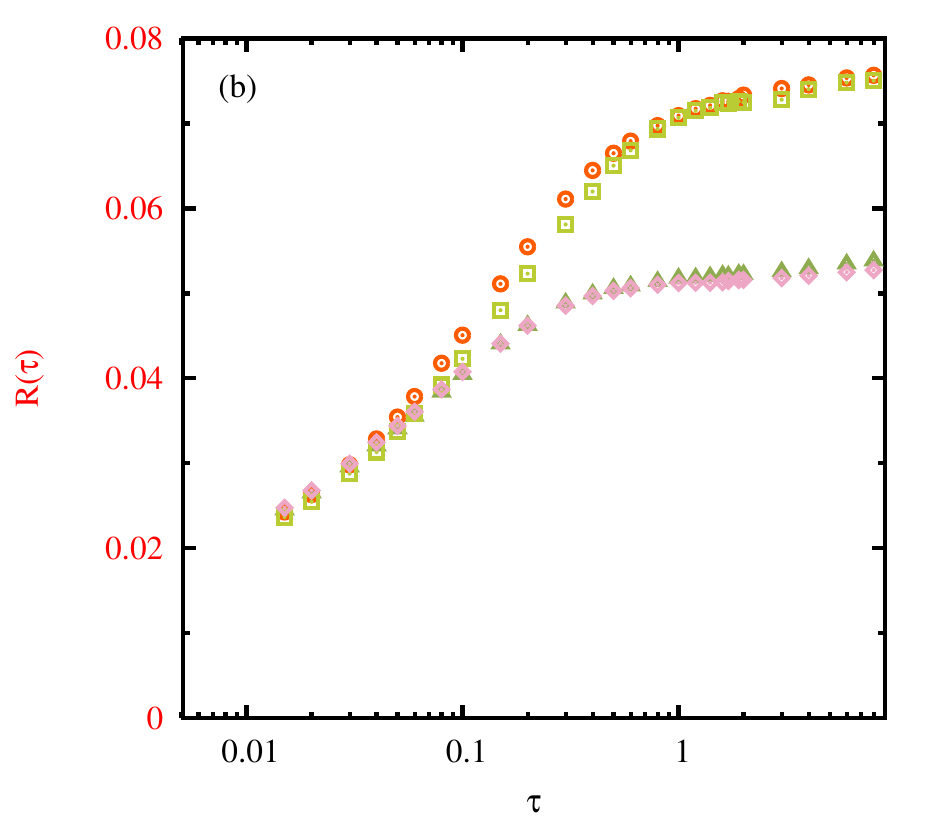}
\includegraphics[height=4.7cm]{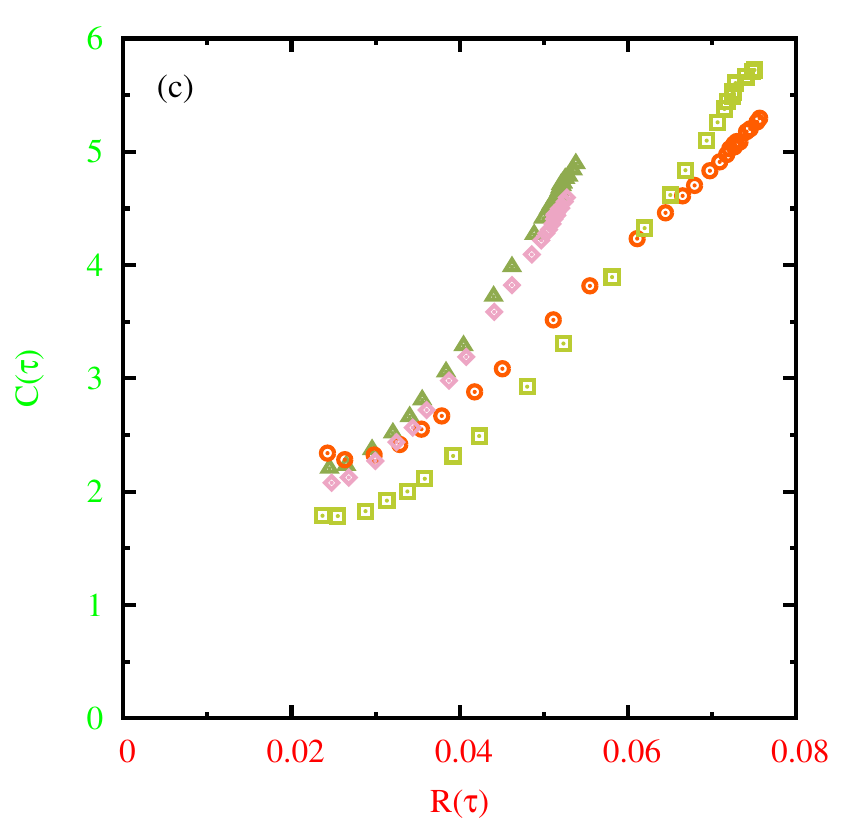}

\caption{\label{cfigExperimental}
Experimental measurement of
(а): $C(\tau)$  (\ref{CCurrentC}),
(b): $R(\tau)$ (\ref{RCurrentR:Dasha:Invariant}),
and (c): $[R(\tau),C(\tau)]$ parametric plot
with a clear linear dependence observed
(similar to
Fig. \figureRef{cfigCircuitModellingCapacitance}{b} modeling).
The results are presented for
    four commercial supercapacitors
    \CapBlueLink{} (circles),
    \CapGreenLink{} (triangles),
    \CapBlackLink{} (squares),
    and \CapWhiteLink{} (rhombuses).
    They have $C/R$ characteristic slope of
    $65 F/\Omega$, $111 F/\Omega$, $92 F/\Omega$ and $107 F/\Omega$  respectively.
}
\end{figure}

The modeling of previous section shows the value
of developed technique. Consider its practical application.
In experiments we tested the approach on four commercial supercapacitors,
see Fig. \ref{cfigExperimental} for the list,
with the initial potential $U_0=2.5V$; these supercapacitors all have $5F$ nominal capacitance
and are $2.7V$ rated. Shorting time duration $\tau$
was taken $10^{-2}$ to $10^2~\mathrm{s}$, the $[10^{-2}\div 10]~\mathrm{s}$
is an informative interval as the plateau is reached at $\tau>10~\mathrm{s}$.

To obtain $[R(\tau),C(\tau)]$
we need to know the integrals $\int Idt$ (\ref{CCurrentC}) and $\int I^2dt$ (\ref{RCurrentR:Dasha:Invariant})
of current, they correspond to total charge and dissipated energy respectively.
The measurement is implemented with
\href{https://www.st.com/en/microcontrollers-microprocessors/stm32-32-bit-arm-cortex-mcus.html}{STM32F103C8T6 ARM}
microcontroller.
Operational amplifier \href{https://www.analog.com/media/en/technical-documentation/data-sheets/AD823.pdf}{AD823} brings small potential on shorting stage to
the range of maximal ADC precision.
We calculate current moments by direct integration:
\begin{align}
Q(\tau)=\int\limits_0^{\tau} I dt&\approx\sum\limits_{k} \frac{U(t_k)}{R_s}(t_k-t_{k-1})
\label{Qdc} \\
 \int\limits_0^{\tau} I^2 dt&\approx\sum\limits_{k} \frac{U^2(t_k)}{R^2_s}(t_k-t_{k-1})
 \label{I2dc}
\end{align}
``Right rectangle'' integration rule is used to simplify
microcontroller implementation, it is more than adequate for
a typical sampling frequency $10^5/\mathrm{s}$.
Previously considered\cite{kompan2021inverse} minimal internal resistance
(\ref{RiCurrentC}) requires only a jump in potential.
Similar current-‐interruption technique is often used
in fuel cell measurements \cite{larminie2003fuel}, page 64,
the immediate rise voltage $V=IR_i$ is an analogue of $U_1-U^*$;
the \cite{larminie2003fuel} technique is equivalent to Eq. (\ref{RiCurrentC}), where current
interruption from $I_1$ to $0$ gives immediate rise
(initial rebound) $R_1I_1=U_1-U^*$
of the potential
what allows to determine the minimal resistance $R_1$
\footnote{
The minimal resistance
corresponds to $\omega\to\infty$ (or $\tau\to 0$) limit.
In transmission line model in Fig. \figureRef{cfigCircuit}{а}
it is the $R_1$ in the circuit. In superposition model
in Fig. \figureRef{cfigCircuit}{b} it is the $1\Big/\sum\limits_k \frac{1}{R_k}$.
}(and only the $R_1$).
Eq. (\ref{RCurrentR:Dasha:Invariant}) has a major advantage over
this current interruption technique.
It uses second order moment of current (\ref{I2dc})
and the $R(\tau)$, which contains information about internal $RC$
distribution, is obtained.
Another advantage is that (\ref{I2dc}) calculates an integral over
the entire shorting interval,
what makes it less measurement error prone compared to the measurement of
immediate rise voltage that is a single point observation
what can possibly give some discrepancy in practical measurements. With accurate
measurement (confirmed by the modeling) we always have $R_1=R(\tau\to 0)$,
the minimal possible $R(\tau)$.

Experimental results are similar to Figs. \ref{cfigCircuitModellingCapacitance}
and \ref{cfigCircuitModellingNetwork}
modeling.
Both $R(\tau)$ and $C(\tau)$ grow with $\tau$ increase.
The most informative is a parametric plot $[R(\tau),C(\tau)]$
in Fig. \figureRef{cfigExperimental}{c}. With $\tau$ above some value the
$[R(\tau),C(\tau)]$ is almost linear function, even more linear than
three $RC$ transmission line model in
Fig. \figureRef{cfigCircuitModellingCapacitance}{b},
a convexity  at low $\tau$ make the plot similar
to binary tree model in Fig. \figureRef{cfigCircuitModellingNetwork}{e,f}.
Initial $R$-offset is determined by the contacts.
The slope $\Delta C/\Delta R$ is the most important characteristic
determining possible power properties of the device
in terms of materials and technology used.
We call it $C/R$ characteristic slope (measured in $F/\Omega$):
how much $\Delta C$ we can ``gain'' if
we are ready to ``lose'' $\Delta R$ in internal resistance.
From the data in Fig. \figureRef{cfigExperimental}{c} it immediately follows
that the maximal characteristic slope is
$111 F/\Omega$ for \CapGreenLink{} (triangles) and
the minimal characteristic slope is $65 F/\Omega$
for \CapBlueLink{} (circles). This linear $C/R$ dependence,
measured completely in time domain is
the major result of this work. The
$C/R$ characteristic slope
can be viewed as a measure of supercapacitor perfection.

A SC, when probed at different time scales,
has charge penetration to pores increasing with time scale.
The equivalent capacitance growths.
But the equivalent internal
resistance also growths.
The perfection is considered as a relative
contribution of deep pores to capacitance and to internal resistance.
 The more deep pores contribute
to capacitance and the less to resistance --- the better supercapacitor is.
Their relative contribution is determined by the slope in $[R(\tau),C(\tau)]$
parametric plot.

\section{\label{frequencyDomainExperiments}Supercapacitors Experimental Measurements in Frequency Domain}
\begin{figure}

\includegraphics[height=6.7cm]{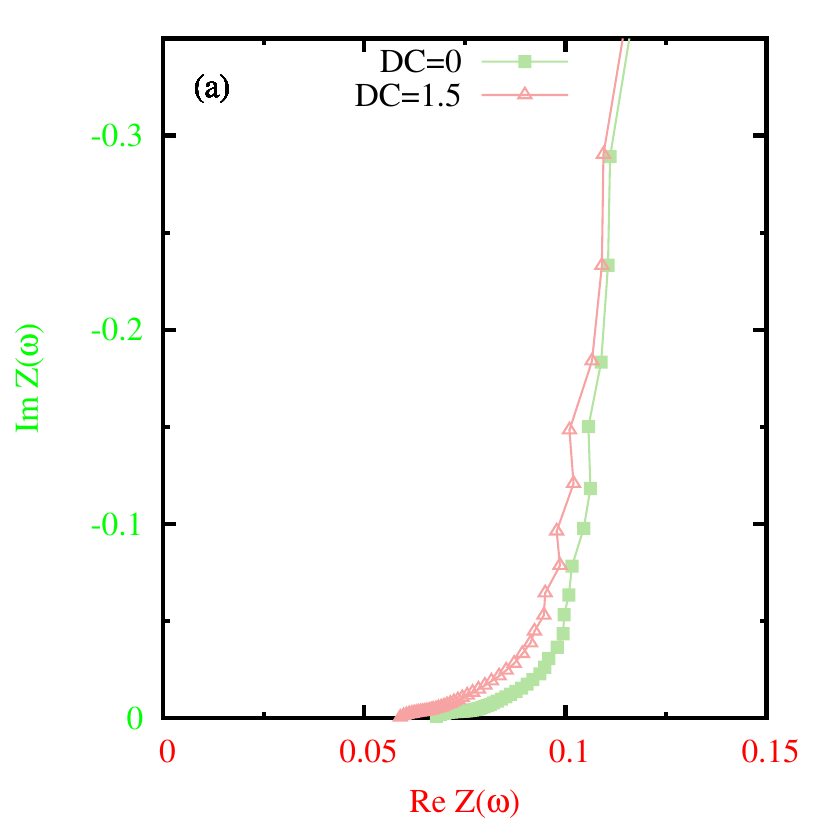}
\includegraphics[height=6.7cm]{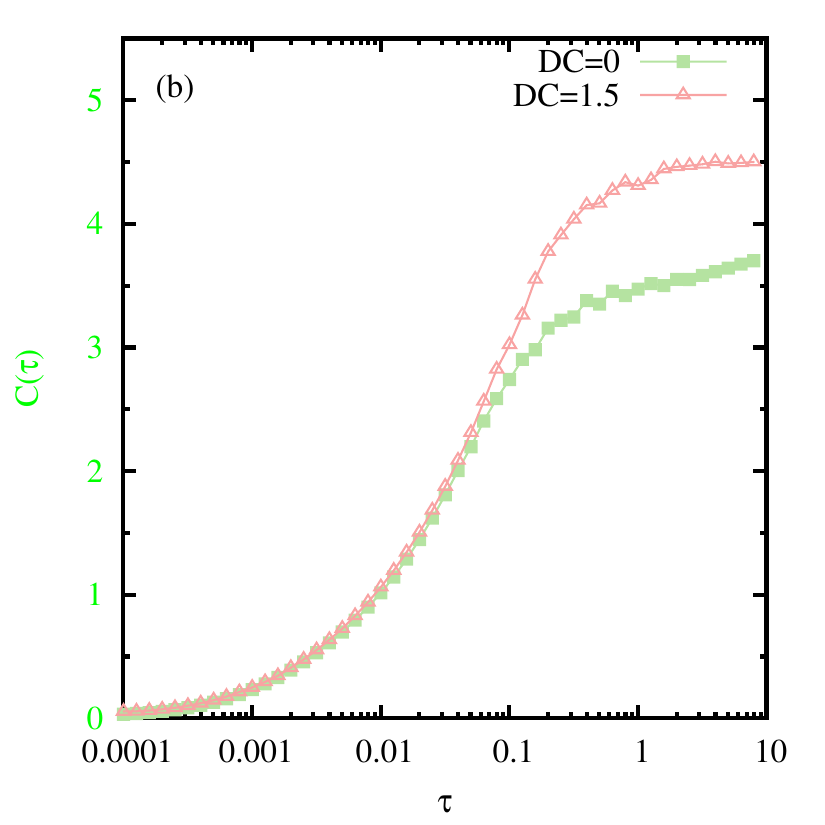} \\
\includegraphics[height=6.7cm]{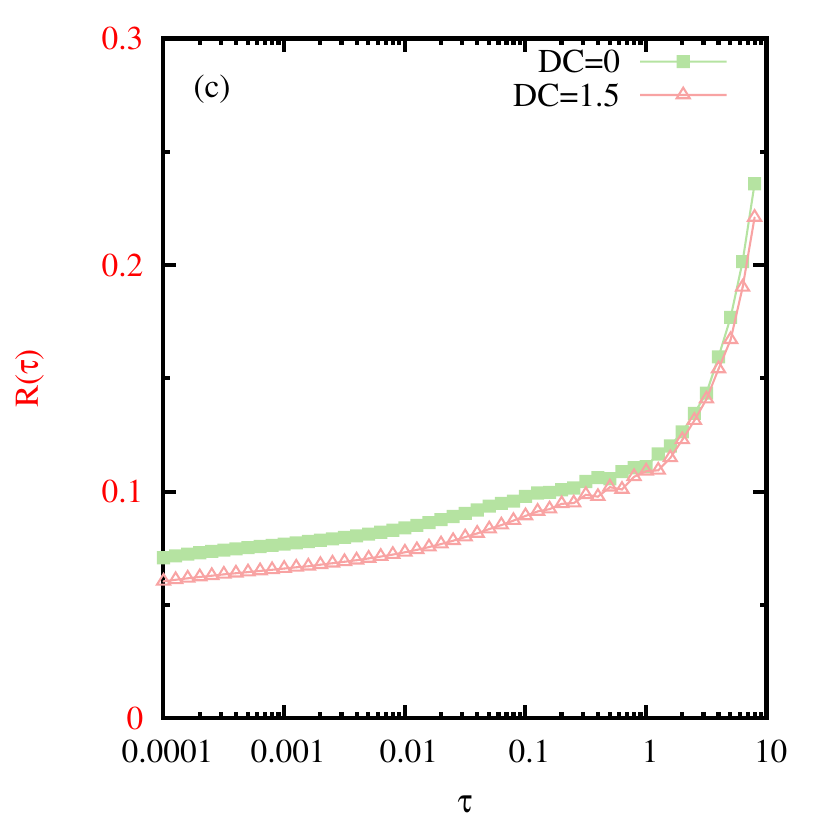}
\includegraphics[height=6.7cm]{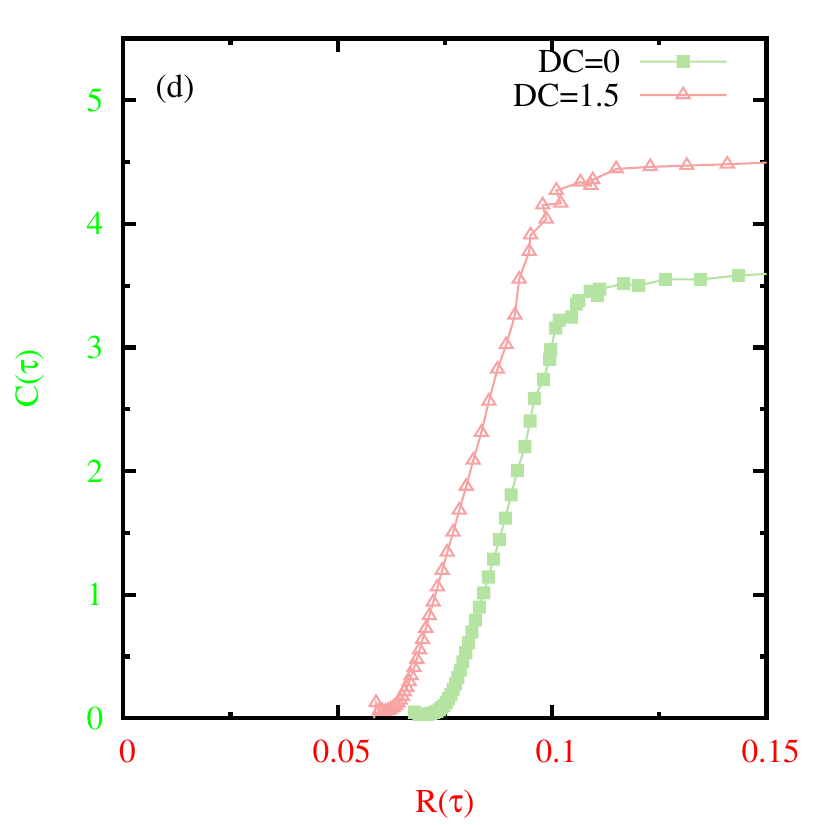}
\caption{\label{impedanceFig}
For  \CapBlackLink{} supercapacitor.
(a) Original Nyquist plot.
(b) $C(\tau)$ dependence (\ref{CimpEq}).
(c) $R(\tau)$ dependence (\ref{RimpEq}).
(d) The $[R(\tau),C(\tau)]$ parametric plot. A linear dependence
is clearly observed within two orders of $\tau$. Compare with
Fig. \figureRef{cfigExperimental}{c}.
}
\end{figure}

Obtained in previous section time--domain $C/R$ linear dependence
is an important new result.
From methodological point of view, to prove the technique,
we consider the same $C/R$ dependence in frequency domain.
As we \hyperref[parallelCircuit]{emphasized} above
the Eqs. (\ref{CimpEq}) and (\ref{RimpEq}) are not accurate
for real supercapacitors, however in this section we apply
them to supercapacitors assuming no charge leak.
Impedance data is measured in frequency range $10^{-2}\div 10^3~\mathrm{Hz}$
with $10mV$ AC amplitude.
Nyquist plot for \CapBlackLink{}
is presented in Fig. \figureRef{impedanceFig}{a}. The impedance
was measured with two values of bias applied: $0V$ and $1.5V$;
it is typical for supercapacitors to have the parameters slightly changed
under bias applied.\footnote{
With $U$-dependent properties one can make Nyquist parametric plot
at fixed frequency $\omega=const$ with $U$ being the parameter
$[\mathrm{Re} Z(U),\mathrm{Im} Z(U)]$
 to obtain a vertical half-circle, see  Fig. \figureRef{differentRCcurcuits}{c}
 and \cite{kompan2021impedanceHodograph}.
}
In Fig. \figureRef{impedanceFig}{b,c}
the $C(\tau)$ (\ref{CimpEq}) and $R(\tau)$ (\ref{RimpEq}) are presented.
The $R(\tau)$ is valid only for $\tau\lesssim 1~\mathrm{s}$, then it diverges.
The divergence is caused both by limited applicability of
Eqs. (\ref{CimpEq}), (\ref{RimpEq})
and increased measurement errors at low frequencies.
The $C(\tau)$ is valid for $\tau\gtrsim 10^{-2}~\mathrm{s}$,
due to limited applicability of
Eq. (\ref{CimpEq}).

\begin{figure}
\includegraphics[height=6.7cm]{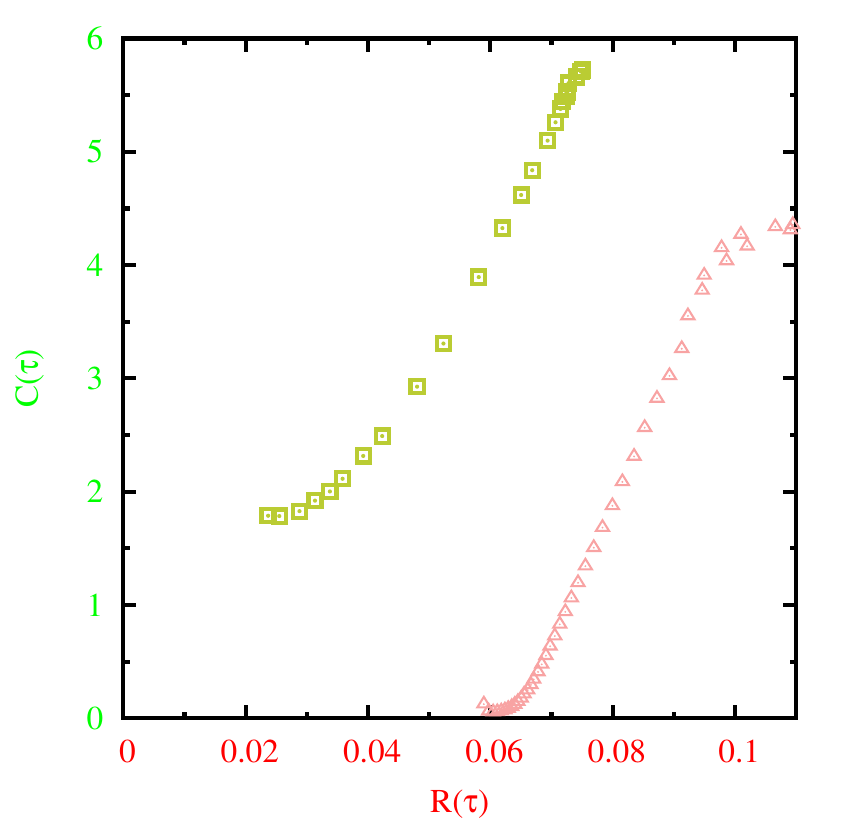}
\caption{\label{compareTW}A comparison of
$C/R$ characteristic slope for \CapBlackLink{}
measured in time domain Fig. \figureRef{cfigExperimental}{c} (squares)
and frequency domain Fig. \figureRef{impedanceFig}{d} (triangles). 
A linear dependence is clearly observed in both cases.
See Fig. \figureRef{cfigCircuitModellingCapacitance}{b}
and Fig. \figureRef{cfigCircuitModellingNetwork}{b,f}
for modeled systems.
}
\end{figure}

However, if we do $[R(\tau),C(\tau)]$ parametric plot,
see Fig. \figureRef{impedanceFig}{d},
we observe a linear dependence in the range $10^{-2}\le\tau\le 1$, similar to the one
in Fig. \figureRef{cfigExperimental}{c} measured in time domain.
The $C/R$ characteristic slope is about $110 F/\Omega$;
this is similar to $92 F/\Omega$ value in time domain.
A comparison
is presented in Fig. \ref{compareTW}. The $C/R$ characteristic slope
is almost the same;
the absolute values are shifted due to \hyperref[parallelCircuit]{limitation}
of Eqs. (\ref{CimpEq}) and (\ref{RimpEq})
frequency domain estimation.
Close to linear $[R(\omega),C(\omega)]$ parametric plot
was observed (in frequency domain) experimentally by other researchers,
see Fig. 2c of \cite{allagui2016spectral}
where it is about $0.013 F/\Omega$ for NEC supercapacitor
part \#FGR0H105ZF, $1F$, $5.5V$ rated (two SC are connected in serial,
the $C/R$ slope is four times lower than for a single SC).
Here we observe a linear  $[R,C]$ parametric plot both
in time- and frequency- domains for a number of different SC.

This leads us to conclude that linear dependence
in parametric plot 
$[R(\tau),C(\tau)]$
within several orders in $\tau$
is a very general property of distributed $RC$ systems,
it can be observed both in time and frequency domains.
It measures supercapacitor perfection.

\section{\label{Discussion}Discussion}

In this work a novel time-domain measurement technique for supercapacitors
characterization is developed, modeled numerically,
and experimental testing
on a number of commercial supercapacitors
was conducted to validate it.
The technique consists in shorting a supercapacitor
on $\tau$ duration and measuring first $\int Idt$ and second $\int I^2dt$
moments of current along with the potential before and after shorting.
The effective $C(\tau)$ and $R(\tau)$ are then obtained
from charge preservation and energy dissipation invariants.
The approach can be considered as an alternative/extension to
commonly used \cite{maxwellSC} DC-ESR
technique, see Maxwell
\href{https://maxwell.com/wp-content/uploads/2021/08/2_7_5F_ds_3001974_datasheet.pdf}{datasheet}; we have an effective resistance $R(\tau)$
calculated at different values of $\tau$.

Among new results obtained with the developed time-domain technique is an observation
of linear behavior in $[R(\tau),C(\tau)]$ parametric plot
that characterizes the device
in terms of materials and technology used.
The $C/R$ characteristic slope
is a constant within several orders of $\tau$,
this has been confirmed in Section \ref{modelingBranch} modeling
on a number of circuit types: transmission line, binary tree, etc.
The result
is also confirmed with impedance technique;
whereas in impedance spectroscopy typically only conformal mapping (such as impedance/admittance $Y=1/Z$, see Fig. \ref{differentRCcurcuits})
are considered, we have shown that non-conformal mapping
from $[\mathrm{Re} Z(\omega),\mathrm{Im} Z(\omega)]$
to $[R(\tau),C(\tau)]$
allows to obtain the $C/R$ characteristic slope
(however time-domain measurement is more accurate
as no conversion required).
A linear dependence in $[R(\omega),C(\omega)]$ parametric plot
in frequency domain 
was experimentally observed in \cite{allagui2016spectral},
this confirms our results.
The $[R(\tau),C(\tau)]$ when directly measured in time domain
gives
the most accurate results (compared to frequency domain measurement)
as it is based on charge preservation and energy dissipation invariants.

\appendix

\section{\label{psiXSoftware}Software Modeling}
The system was modeled in
\href{http://ngspice.sourceforge.net/}{Ngspice circuit simulator}.
The circuit
was created in  gschem  program of
\href{http://www.geda-project.org/}{gEDA} project.
This is an
\href{http://www.ioffe.ru/LNEPS/malyshkin/RCcircuit_ver2.zip}{updated version}
of the original \cite{kompan2021inverse} code.
The newest software version is available at \cite{RCsimulator}.
The differences from the
\href{http://www.ioffe.ru/LNEPS/malyshkin/RCcircuit.zip}{previous version}
are:
\begin{itemize}
\item
\texttt{\seqsplit{extract\_voltages\_and\_current.pl}}
now calculates not only the fist $\int Idt$ but also
the second $\int I^2dt$ moment of current. An identification of potential jump has been improved; currently it checks for absolute or relative jump in the potential.
\item
\texttt{\seqsplit{cmd\_withIL.sh}} script was modified to
automatically run $\tau\in[10^{-4} \div 10^{4}]~\mathrm{s}$.
It now runs on five input files:
\texttt{\seqsplit{Farades\_y\_with\_variables\_current\_save.sch}},
\texttt{\seqsplit{BinaryTree255.net}},
\texttt{\seqsplit{BinaryTree.net}},
\texttt{\seqsplit{Superposition.net}},
\texttt{\seqsplit{Chain.net}}.
Modeling results are presented in  Figs.
\ref{cfigCircuitModellingCapacitance} and \ref{cfigCircuitModellingNetwork}.
The file \texttt{\seqsplit{Farades\_y\_with\_variables\_current\_save.sch}}
corresponds to a three $RC$ transmission line system,
the other files contain a large $RC$ network, they
are generated automatically by running the commands:
\texttt{\seqsplit{java\ com/polytechnik/echem/ChainPrint}} (transmission line of $31$ random $RC$ elements),
\texttt{\seqsplit{java\ com/polytechnik/echem/SuperpositionPrint}} (superposition model of $31$ random $RC$ elements),
and
\texttt{\seqsplit{java\ com/polytechnik/echem/BinaryTreePrint}} (two binary trees with depth equals to $7$ and $4$, $255$ and $31$ random $RC$ elements respectively,
the calculations are performed with depth-first
\href{https://en.wikipedia.org/wiki/Tree_traversal}{recursive tree traversal}).
Each of these commands
creates $Z(\omega)$ impedance file
and corresponding Ngspice
\href{https://ngspice.sourceforge.io/docs/ngspice-manual.pdf}{.net}
file for circuit simulation.
\href{https://commons.apache.org/proper/commons-math/}{Apache math}
library
needs to be installed (\href{https://commons.apache.org/proper/commons-math/userguide/complex.html}{Complex} class is used).
\end{itemize}
Required software to be installed: \verb+perl+, \verb+java+, and \verb+ngspice+.
Run shell script \verb+cmd_withIL.sh+ to model all these $RC$ system
with $\tau$ in $[10^{-4} \div 10^{4}]$ range.

\section{\label{SymCalc}Symbolic verification of $2C/R$ slope
for transmission line model}
The correctness of Eq. (\ref{CRw})
can be verified using
\href{https://docs.sympy.org/latest/index.html}{SymPy}
symbolic calculations library\cite{meurer2017sympy}.
Using $Z(\omega)$ expression (\ref{ZChainValue})
we obtain:
\begin{verbatim}
from sympy import *
w=Symbol('w',real=True)
R=Symbol('R',real=True)
C=Symbol('C',real=True)
Z=Symbol('Z',complex=True)
Z= R/2 + sqrt(R*R/4+R/(I*w*C))
checkLinear=simplify(simplify(
    -1/(w*im(Z)) - 2*C/R*re(Z)
))
print("|checkLinear=",checkLinear)
C_R=simplify(simplify(
    diff(-1/(w*im(Z)),w) / diff(re(Z),w)
))
print("|C_R=",C_R)
\end{verbatim}
Obtained formula 
\texttt{\seqsplit{C\_R=4*(C**2*R**2*w**2*sin(atan2(-R/(C*w), R**2/4)/2) + 2*C*R*w*cos(atan2(-R/(C*w), R**2/4)/2) + 8*sin(atan2(-R/(C*w), R**2/4)/2))/(w*(R**2 - sqrt(1/(C**2*w**2))*sqrt(C**2*R**2*w**2 + 16)*abs(R))*(C*R*w*sin(atan2(-R/(C*w), R**2/4)/2) + 4*cos(atan2(-R/(C*w), R**2/4)/2)))}}
is equal to $2C/R$ constant of Eq. (\ref{CRw}).
This can be easily proven with symbolic
computations, see \texttt{\seqsplit{checkLinear=-1/(w*im(Z)) - 2*C/R*re(Z)}}
which is equal exactly to the constant \texttt{\seqsplit{-C}},
i.e. for an infinite transmission line we have a linear plot
$C_{impedance}(\omega)=\frac{2C}{R}R_{impedance}(\omega)-C$, Eq. (\ref{TLinfLength}).
Alternatively 
see \texttt{\seqsplit{test\_CR\_ratio\_transmission\_line.py}}
that verifies \verb+C_R+ expression by evaluating the
\texttt{\seqsplit{(C\_R-2*C/R)}} at different \verb+R+, \verb+C+, and \verb+w+.

The $C/R$ slope for nTE,
an infinite tree
of identical $RC$ elements
with the number of descendant nodes \verb+n+,
is obtained analytically using $Z(\omega)$ from (\ref{Zsol}).
Put \texttt{\seqsplit{n=0.5}}
 (or \texttt{\seqsplit{n=Symbol('n',real=True)}} in you
 need a symbolic formula),
 and \texttt{\seqsplit{Z= (R+(1-n)/(I*w*C))/2 + sqrt((R+(1-n)/(I*w*C))**2/4+n*R/(I*w*C))}}
into the code above.
The $C/R$ slope for nTE is a rather long formula which is not a constant.
Explicit $Z(\omega)$ is:
the real part
\texttt{\seqsplit{Re(Z)= R/2 + (1/(C**6*w**6))**(1/4)*(4*C**2*R**2*w**2*(n + 1)**2 + (C**2*R**2*w**2 - (n - 1)**2)**2)**(1/4)*cos(atan2(-R*(n + 1)/(2*C*w), (C**2*R**2*w**2 - (n - 1)**2)/(4*C**2*w**2))/2)*Abs(sqrt(C)*sqrt(w))/2}}
and
the imaginary part
\texttt{\seqsplit{Im(Z)= (C*w*(1/(C**6*w**6))**(1/4)*(4*C**2*R**2*w**2*(n + 1)**2 + (C**2*R**2*w**2 - (n - 1)**2)**2)**(1/4)*sin(atan2(-R*(n + 1)/(2*C*w), (C**2*R**2*w**2 - (n - 1)**2)/(4*C**2*w**2))/2)*Abs(sqrt(C)*sqrt(w)) + n - 1)/(2*C*w)}}.
One can convert these SymPy formulas to \LaTeX{} ones
by using \texttt{\seqsplit{print\_latex()}},
however long generated formulas may not fit a single line.
For a general infinite tree
symbolic computations do not give much insight compared to
regular numerical estimation of $Z(\omega)$ from (\ref{Zsol}).
The command ``\texttt{\seqsplit{python3\ n-tree\_element.py\ 0.5}}''
prints the $C/R$ slope formula (\ref{CRslopeW}) expanded and
 for
$\omega\in[10^{-5}\div 10^4]~\mathrm{Hz}$
outputs
$\omega$,
$R_{impedance}(\omega)$,
$C_{impedance}(\omega)$,
$\mathrm{Im} Z(\omega)$,
and  $C/R$ slope for a given $n=0.5$.

\section{\label{AppndDifferentRCexponent}
Self-similar $RC$ networks and nTE model}

Carbon structures of SC electrodes
often have different exponent for $C$ and $R$.
In Fig. \figureRef{cfigCircuitModellingNetwork}{g,h} modeling
we used
different for $C$ and $R$ depth-dependent factors (\ref{betaFact})
to construct a more
realistic RC distribution. Similar approach
is used to construct a self-similar $RC$ network.
Assume we want to calculate $Z(\omega)$ of a transmission
line in Fig. \figureRef{cfigCircuit}{a} with
\begin{subequations}
\label{RCdep}
\begin{align}
R_k&=n_R^kR \label{nRexp} \\
C_k&=n_C^kC \label{nCexp}
\end{align}
\end{subequations}
Two exponents $n_R$ and $n_C$ make it more difficult to study.
Regular nTE analytic solution (\ref{Zsol})
corresponds to $n=n_C=1/n_R$.
Recurrence relation in general case is
\begin{align}
Z_{k}&=R_k+\frac{Z_{k+1}\frac{1}{j\omega C_k}}
{Z_{k+1}+\frac{1}{j\omega C_k}}
\label{recurrZc}
\end{align}
When applied to $RC$ network (starting with the largest $k$)
we obtain
a composition of 
\href{https://en.wikipedia.org/wiki/Generalized_continued_fraction#Linear_fractional_transformations}{linear fractional transformations}
(M\"{o}bius transformation):
\begin{align}
T_k&=\left(
\begin{matrix}
R_k+\frac{1}{j\omega C_k} & \frac{R_k}{j\omega C_k} \\
1& \frac{1}{j\omega C_k}
\end{matrix}
\right)
\label{TMatrix} \\
\mathcal{T}&=
T_1T_2\dots T_{k-1}T_k \label{TZtransformProd} \\
Z&=\frac{\mathcal{T}_{00}Z_{k+1}+\mathcal{T}_{01}}{\mathcal{T}_{10}Z_{k+1}+\mathcal{T}_{11}}
\label{ZfromT}
\end{align}
with the determinant $-1/\left(\omega C_k\right)^2$.
Transformation matrix corresponding to two $RC$ elements
is equal to matrix product (\ref{TZtransformProd})
of two individual transformations.
In $k\to\infty$ limit
with $Z_{k\to\infty}=\infty$ boundary condition
the  $Z(\omega)$ is
a transcendental function.\footnote{
Note that for $k$-independent $R_k$ and $C_k$
the $Z(\omega)$ solutions (\ref{ZChainValue})
correspond to the fixed points of a M\"{o}bius transformation (\ref{ZfromT})
with $\mathcal{T}=T_k$.}
At fixed $k$ it is a ratio of two $k$-degree polynomials on $\omega$ ---
the ratio of $\mathcal{T}_{00}$ and $\mathcal{T}_{01}$ elements
of combined transformation matrix (\ref{TZtransformProd}).
There is no
an analytic solution to $Z(\omega)$ in general $n_R$, $n_C$ case (\ref{RCdep}), but
it is easy to calculate
numerically. The command
 ``\texttt{\seqsplit{java\ com/polytechnik/echem/TransmissionLineGeometricProgression\ 200\ 1.5\ 0.9}}''
builds $200$ elements transmission line (\ref{RCdep})
with $n_R=1.5$ and $n_C=0.9$,
then it calculates $Z(\omega)$.
A remarkable feature of self-similar $RC$ models --
they can be used to model CPE (\ref{CPEZw}), see \cite{dutta2015resistive,kartci2020history}
for a review.
If $n_R\ge 1$ and $n_C\ge n_R$ then (\ref{RCdep})
$RC$ network has $Z(\omega)$ that is similar to CPE
with $0\le\alpha\le 0.5$.
If $n_C\ge1$ and $n_R\ge n_C$ then (\ref{RCdep})
$RC$ network  has $Z(\omega)$ that is similar to CPE
with $0.5\le\alpha\le 1$. 

The dependence of $\alpha$ of CPE modeled by $RC$-network (\ref{RCdep})
on $n_C$ and $n_R$
is non-linear and is a subject of future research,
the value of $\alpha$  increases with $n_R$ and decreases with $n_C$.
Both $n_R$ and $n_C$ must be greater than $1$, otherwise
the system behavior is very different from CPE.
We tried to use the expressions
$R_{k+1}/R_k=\mathscr{R}^{\alpha}$
and
$C_k/C_{k+1}=\mathscr{R}^{1-\alpha}$ from Eq. (11)
or $C_{k+1}/C_{k}=\mathscr{R}^{1-\alpha}$ from Eq. (9)
of Ref. \cite{sugi2002frequency}. 
With them $\alpha$ either does not depend on $n_R/n_C$
or is a function on $n_R/n_C$ only ---
this contradicts our numerical experiments.
For a good linear behavior 
both $n_R$ and $n_C$ should be greater than $1$, a good choice for
the lowest one is about $1$ (e.g. for $\alpha\le 0.5$ put $n_R=1$
(or $n_R=1.02$) then
find $n_C\ge n_R$ producing required value of $\alpha$; 
for $\alpha\ge 0.5$ put $n_C=1$
(or $n_C=1.02$) then
find $n_R\ge n_C$ producing required value of $\alpha$).

Similarly to the transmission line
model,
the scaling (\ref{RCdep}) also gives a good CPE--like behavior\cite{valsa2013rc}
for superposition model in Fig.  \figureRef{cfigCircuit}{b}.
This model corresponds to
\begin{align}
T_k&=\left(
\begin{matrix}
R_k+\frac{1}{j\omega C_k} & 0 \\
1& R_k+\frac{1}{j\omega C_k}
\end{matrix}
\right)
\label{TMatrixSup}
\end{align}
linear fractional transformations.
The $\alpha$ increases with $n_R$ and decreases with $n_C$.
With $n_R\le1$ the model produces capacitance-like behavior,
only at $n_R>1$ it has CPE-like behavior.
The values $n_R<1$, $n_C<1$ allow to build
a CPE element at high frequencies\cite{valsa2013rc}.
As we study supercapacitors we are most interested
in low-frequency behavior.\footnote{
For superposition model one can possibly take
$\widetilde{C}=C_{k_{\max}}$ and re-numerate all $C$
with $\widetilde{n_C}=1/n_C$ factor.
This maps $n_C>1$ solution to a one with $n_C<1$.
This is not possible,
however, for transmission line model.
}
A CPE-like behavior
in a wide range of \textsl{low frequencies} can be obtained
 with $n_R>1$, $n_C>1$.
For example try $n_R=2$, $n_C=3$ to obtain $\alpha<0.5$, and
$n_R=2$, $n_C=1.5$ to obtain $\alpha>0.5$ examples.
 The command
 ``\texttt{\seqsplit{java\ com/polytechnik/echem/SuperpositionGeometricProgression\ 100\ 2\ 1.5}}''
builds $100$ $RC$ elements superposition model
in Fig. \figureRef{cfigCircuit}{b} having (\ref{RCdep}) elements
with $n_R=2$ and $n_C=1.5$,
then it calculates $Z(\omega)$. For
$1<n_R=n_C\lesssim 2.5$ the superposition model produces a good CPE with $\alpha=0.5$.

In Fig. \figureRef{cfigCircuitModellingNetwork}{l}
we present CPE modeled by
Fig. \figureRef{cfigCircuit}{a}
transmission line
containing
$100$
self-similar
$RC$ elements (\ref{RCdep})
 with various $n_R$ and $n_C$.
Besides already considered $\alpha=0.5$ nTE case $n=1$,
we see a very close to linear
$[\mathrm{Re} Z(\omega),\mathrm{Im} Z(\omega)]$
behavior, especially for $\alpha>0.5$
what corresponds to $n_R\ge n_C$ and $n_C=1$. The plot is
less linear for $\alpha<0.5$ --- this is $n_C\ge n_R$ and $n_R=1$ case.

This simple CPE modeling leads us to conclude that self-similar models
is a good choice to model CPE and other fractional $RC$ networks.
An important advantage of $RC$ network 
is that it can be used for time-domain considerations.
It is very difficult to apply
(\ref{CCurrentC}) and (\ref{RCurrentR:Dasha:Invariant})
to a model, such as CPE (\ref{CPEZw}), that is defined in frequency domain.
A model consisting of actual resistors and capacitors, such as
(\ref{RCdep}),
can be directly studied (either experimentally or modeling) using
time domain spectroscopy
we have developed in this paper.

\bibliographystyle{elsarticle-num.bst}
\bibliography{echem,LD}

\begin{thebibliography}{10}
\expandafter\ifx\csname url\endcsname\relax
  \def\url#1{\texttt{#1}}\fi
\expandafter\ifx\csname urlprefix\endcsname\relax\def\urlprefix{URL }\fi
\expandafter\ifx\csname href\endcsname\relax
  \def\href#1#2{#2} \def\path#1{#1}\fi

\bibitem{yoo2016fast}
Y.~Yoo, M.-S. Kim, J.-K. Kim, Y.~S. Kim, W.~Kim, Fast-response supercapacitors
  with graphitic ordered mesoporous carbons and carbon nanotubes for ac line
  filtering, Journal of Materials Chemistry A 4~(14) (2016) 5062--5068.
\newblock \href {https://doi.org/10.1039/C6TA00921B}
  {\path{doi:10.1039/C6TA00921B}}.

\bibitem{borenstein2017carbon}
A.~Borenstein, O.~Hanna, R.~Attias, S.~Luski, T.~Brousse, D.~Aurbach,
  {Carbon-based composite materials for supercapacitor electrodes: a review},
  Journal of Materials Chemistry A 5~(25) (2017) 12653--12672.
\newblock \href {https://doi.org/10.1039/C7TA00863E}
  {\path{doi:10.1039/C7TA00863E}}.

\bibitem{kompan2019reverse}
M.~E. Kompan, V.~G. Malyshkin, {The Reverse Relaxation Effect and Structure of
  Porous Electrodes in Supercapacitors}, Technical Physics Letters 45~(1)
  (2019) 45--47.
\newblock \href {https://doi.org/10.1134/S1063785019010279}
  {\path{doi:10.1134/S1063785019010279}}.

\bibitem{il2020modeling}
D.~S. Il’yushchenkov, A.~A. Tomasov, S.~A. Gurevich, {Modeling
  Charge/Discharge Characteristics of Supercapacitors on the Basis of an
  Equivalent Scheme with Fixed Parameters}, Technical Physics Letters 46 (2020)
  80--82.
\newblock \href {https://doi.org/10.1134/S1063785020010253}
  {\path{doi:10.1134/S1063785020010253}}.

\bibitem{ghanbari2021self}
T.~Ghanbari, E.~Moshksar, S.~Hamedi, F.~Rezaei, Z.~Hosseini, {Self-discharge
  modeling of supercapacitors using an optimal time-domain based approach},
  Journal of Power Sources 495 (2021) 229787.
\newblock \href {https://doi.org/10.1016/j.jpowsour.2021.229787}
  {\path{doi:10.1016/j.jpowsour.2021.229787}}.

\bibitem{pourkheirollah2023simplified}
H.~Pourkheirollah, J.~Keskinen, M.~M{\"a}ntysalo, D.~Lupo, {Simplified
  exponential equivalent circuit models for prediction of printed
  supercapacitor's discharge behavior-Simulations and experiments}, Journal of
  Power Sources 567 (2023) 232932.
\newblock \href {https://doi.org/10.1016/j.jpowsour.2023.232932}
  {\path{doi:10.1016/j.jpowsour.2023.232932}}.

\bibitem{maxwellSC}
{Maxwell Techonologies}, {BCAP0005 P270 S01, ESHSR-0005C0-002R7,
  \href{https://maxwell.com/wp-content/uploads/2021/08/2_7_5F_ds_3001974_datasheet.pdf}{Document
  3001974-EN.3},
  \href{https://maxwell.com/wp-content/uploads/2021/08/ProductMatrix.pdf}{product
  list}, and
  \href{https://maxwell.com/wp-content/uploads/2021/08/1007239_EN_test_procedures_technote_2.pdf}{Test
  Procedures for Capacitance, ESR, Leakage Current and Self-Discharge
  Characterizations of Ultracapacitors}.} (2021).
\newblock
  \href{https://maxwell.com/wp-content/uploads/2021/08/1007239_EN_test_procedures_technote_2.pdf}{[link]}.
\newline\urlprefix\url{https://maxwell.com/wp-content/uploads/2021/08/1007239_EN_test_procedures_technote_2.pdf}

\bibitem{IEC62391}
\href{https://webstore.iec.ch/publication/23570}{{IEC 62391-1:2015 RLV}},
  {Fixed electric double-layer capacitors for use in electric and electronic
  equipment} (2015).
\newline\urlprefix\url{https://webstore.iec.ch/publication/23570}

\bibitem{bard2022electrochemical}
A.~J. Bard, L.~R. Faulkner, H.~S. White, {Electrochemical methods: fundamentals
  and applications}, John Wiley \& Sons, 2022.

\bibitem{lasia2002electrochemical}
A.~Lasia, Electrochemical impedance spectroscopy and its applications,
  Springer, 2002.
\newblock \href {https://doi.org/10.1007/978-1-4614-8933-7}
  {\path{doi:10.1007/978-1-4614-8933-7}}.

\bibitem{bagotsky2015electrochemical}
V.~S. Bagotsky, A.~M. Skundin, Y.~M. Volfkovich, {Electrochemical power
  sources: batteries, fuel cells, and supercapacitors}, John Wiley \& Sons,
  2015.
\newblock \href {https://doi.org/10.1002/9781118942857}
  {\path{doi:10.1002/9781118942857}}.

\bibitem{cheng2009assessments}
Y.~Cheng, {Assessments of energy capacity and energy losses of supercapacitors
  in fast charging--discharging cycles}, IEEE Transactions on energy conversion
  25~(1) (2009) 253--261.
\newblock \href {https://doi.org/10.1109/TEC.2009.2032619}
  {\path{doi:10.1109/TEC.2009.2032619}}.

\bibitem{yang2020comparative}
H.~Yang, {A comparative study of supercapacitor capacitance characterization
  methods}, Journal of Energy Storage 29 (2020) 101316.
\newblock \href {https://doi.org/10.1016/j.est.2020.101316}
  {\path{doi:10.1016/j.est.2020.101316}}.

\bibitem{allagui2018short}
A.~Allagui, D.~Zhang, A.~S. Elwakil, {Short-term memory in electric
  double-layer capacitors}, Applied Physics Letters 113~(25) (2018) 253901.
\newblock \href {https://doi.org/10.1063/1.5080404}
  {\path{doi:10.1063/1.5080404}}.

\bibitem{zhang2015supercapacitors}
S.~Zhang, N.~Pan, {Supercapacitors performance evaluation}, Advanced Energy
  Materials 5~(6) (2015) 1401401.
\newblock \href {https://doi.org/10.1002/aenm.201401401}
  {\path{doi:10.1002/aenm.201401401}}.

\bibitem{burke2011power}
A.~Burke, M.~Miller, {The power capability of ultracapacitors and lithium
  batteries for electric and hybrid vehicle applications}, Journal of Power
  Sources 196~(1) (2011) 514--522.
\newblock \href {https://doi.org/10.1016/j.jpowsour.2010.06.092}
  {\path{doi:10.1016/j.jpowsour.2010.06.092}}.

\bibitem{allagui2016spectral}
A.~Allagui, A.~S. Elwakil, B.~J. Maundy, T.~J. Freeborn, {Spectral capacitance
  of series and parallel combinations of supercapacitors}, ChemElectroChem
  3~(9) (2016) 1429--1436.
\newblock \href {https://doi.org/10.1002/celc.201600249}
  {\path{doi:10.1002/celc.201600249}}.

\bibitem{baboo2023investigating}
J.~P. Baboo, E.~Jakubczyk, M.~A. Yatoo, M.~Phillips, S.~Grabe, M.~Dent, S.~J.
  Hinder, J.~F. Watts, C.~Lekakou, {Investigating battery-supercapacitor
  material hybrid configurations in energy storage device cycling at 0.1 to 10C
  rate}, Journal of power sources 561 (2023) 232762.
\newblock \href {https://doi.org/10.1016/j.jpowsour.2023.232762}
  {\path{doi:10.1016/j.jpowsour.2023.232762}}.

\bibitem{kompan2021inverse}
M.~E. Kompan, V.~G. Malyshkin, {On the inverse relaxation approach to
  supercapacitors characterization}, Journal of Power Sources 484 (2021)
  229257.
\newblock \href {https://doi.org/10.1016/j.jpowsour.2020.229257}
  {\path{doi:10.1016/j.jpowsour.2020.229257}}.

\bibitem{barsoukov2018impedance}
E.~Barsoukov, J.~R. Macdonald, {Impedance spectroscopy: theory, experiment, and
  applications}, John Wiley \& Sons, 2018.
\newblock \href {https://doi.org/10.1002/9781119381860}
  {\path{doi:10.1002/9781119381860}}.

\bibitem{valsa2013rc}
J.~Valsa, J.~Vlach, {RC models of a constant phase element}, International
  Journal of Circuit Theory and Applications 41~(1) (2013) 59--67.
\newblock \href {https://doi.org/10.1002/cta.785} {\path{doi:10.1002/cta.785}}.

\bibitem{lavrent1973methods}
M.~A. Lavrent’ev, B.~V. Shabat,
  \href{https://urss.ru/cgi-bin/db.pl?lang=Ru&blang=ru&page=Book&id=64427}{{Methods
  of the Theory of Functions of a Complex Variable}} (1973).
\newline\urlprefix\url{https://urss.ru/cgi-bin/db.pl?lang=Ru&blang=ru&page=Book&id=64427}

\bibitem{carrier2005functions}
G.~F. Carrier, M.~Krook, C.~E. Pearson, {Functions of a complex variable:
  theory and technique}, SIAM, 2005.
\newblock \href {https://doi.org/10.1137/1.9780898719116}
  {\path{doi:10.1137/1.9780898719116}}.

\bibitem{kompan2021impedanceHodograph}
M.~E. Kompan, V.~G. Malyshkin, {Impedance Hodograph for the Parallel RC Circuit
  with Alternating Active Resistance}, Russian Journal of Electrochemistry 57
  (2021) 949--952.
\newblock \href {https://doi.org/10.1134/S1023193521080061}
  {\path{doi:10.1134/S1023193521080061}}.

\bibitem{lozano2003influence}
D.~Lozano-Castello, D.~Cazorla-Amor{\'o}s, A.~Linares-Solano, S.~Shiraishi,
  H.~Kurihara, A.~Oya, {Influence of pore structure and surface chemistry on
  electric double layer capacitance in non-aqueous electrolyte}, Carbon 41~(9)
  (2003) 1765--1775.
\newblock \href {https://doi.org/10.1016/S0008-6223(03)00141-6}
  {\path{doi:10.1016/S0008-6223(03)00141-6}}.

\bibitem{fuertes2004influence}
A.~B. Fuertes, F.~Pico, J.~M. Rojo, {Influence of pore structure on electric
  double-layer capacitance of template mesoporous carbons}, Journal of Power
  Sources 133~(2) (2004) 329--336.
\newblock \href {https://doi.org/10.1016/j.jpowsour.2004.02.013}
  {\path{doi:10.1016/j.jpowsour.2004.02.013}}.

\bibitem{fu2011hierarchical}
R.~wen Fu, Z.~hui Li, Y.~ru~Liang, F.~Li, F.~Xu, D.~cai Wu, {Hierarchical
  porous carbons: design, preparation, and performance in energy storage}, New
  Carbon Materials 26~(3) (2011) 171--179.
\newblock \href {https://doi.org/10.1016/S1872-5805(11)60074-7}
  {\path{doi:10.1016/S1872-5805(11)60074-7}}.

\bibitem{wang2022controllable}
X.~Wang, J.~Xu, B.~Hu, N.~Yuan, X.~Cao, F.~Zhang, R.~Zhang, J.~Ding,
  {Controllable adjustment strategies for activated carbon and application in
  supercapacitors with both ultra-high capacitance and rate performance},
  Diamond and Related Materials 130 (2022) 109466.
\newblock \href {https://doi.org/10.1016/j.diamond.2022.109466}
  {\path{doi:10.1016/j.diamond.2022.109466}}.

\bibitem{danielyan2007increasing}
M.~I. Danielyan, K.~S. Kulakov, S.~L. Kulakov, V.~L. Tumanov, M.~E. Kompan,
  {Increasing the efficiency of metal--air current sources operating in a
  pulse-train mode}, Technical Physics Letters 33~(7) (2007) 597--599.
\newblock \href {https://doi.org/10.1134/S1063785007070176}
  {\path{doi:10.1134/S1063785007070176}}.

\bibitem{kompan2010nonlinear}
M.~E. Kompan, V.~P. Kuznetsov, V.~G. Malyshkin, {Nonlinear impedance of
  solid-state energy-storage ionisters}, Technical Physics 55~(5) (2010)
  692--698.
\newblock \href {https://doi.org/10.1134/S1063784210050142}
  {\path{doi:10.1134/S1063784210050142}}.

\bibitem{ArxivMalyshkinLebesgue}
V.~G. Malyshkin, \href{https://arxiv.org/abs/1807.06007}{{On Lebesgue Integral
  Quadrature}}, ArXiv e-prints (Jul. 2018).
\newblock \href {http://arxiv.org/abs/1807.06007} {\path{arXiv:1807.06007}},
  \href {https://doi.org/10.48550/arXiv.1807.06007}
  {\path{doi:10.48550/arXiv.1807.06007}}.
\newline\urlprefix\url{https://arxiv.org/abs/1807.06007}

\bibitem{ivanovNenashevAleshin2022low}
A.~M. Ivanov, G.~V. Nenashev, A.~N. Aleshin, {Low-frequency noise and impedance
  spectroscopy of device structures based on perovskite-graphene oxide
  composite films}, Journal of Materials Science: Materials in Electronics
  33~(27) (2022) 21666--21676.
\newblock \href {https://doi.org/10.1007/s10854-022-08955-7}
  {\path{doi:10.1007/s10854-022-08955-7}}.

\bibitem{fletcher2014universal}
S.~Fletcher, V.~J. Black, I.~Kirkpatrick, {A universal equivalent circuit for
  carbon-based supercapacitors}, Journal of Solid State Electrochemistry 18
  (2014) 1377--1387.
\newblock \href {https://doi.org/10.1007/s10008-013-2328-4}
  {\path{doi:10.1007/s10008-013-2328-4}}.

\bibitem{devillers2014review}
N.~Devillers, S.~Jemei, M.-C. P{\'e}ra, D.~Bienaim{\'e}, F.~Gustin, {Review of
  characterization methods for supercapacitor modelling}, Journal of Power
  Sources 246 (2014) 596--608.
\newblock \href {https://doi.org/10.1016/j.jpowsour.2013.07.116}
  {\path{doi:10.1016/j.jpowsour.2013.07.116}}.

\bibitem{logerais2015modeling}
P.-O. Logerais, M.~Camara, O.~Riou, A.~Djellad, A.~Omeiri, F.~Delaleux,
  J.~Durastanti, {Modeling of a supercapacitor with a multibranch circuit},
  international journal of hydrogen energy 40~(39) (2015) 13725--13736.
\newblock \href {https://doi.org/10.1016/j.ijhydene.2015.06.037}
  {\path{doi:10.1016/j.ijhydene.2015.06.037}}.

\bibitem{zhang2018review}
L.~Zhang, X.~Hu, Z.~Wang, F.~Sun, D.~G. Dorrell, {A review of supercapacitor
  modeling, estimation, and applications: A control/management perspective},
  Renewable and Sustainable Energy Reviews 81 (2018) 1868--1878.
\newblock \href {https://doi.org/10.1016/j.rser.2017.05.283}
  {\path{doi:10.1016/j.rser.2017.05.283}}.

\bibitem{sen2018implicit}
M.~Sen, J.~P. Hollkamp, F.~Semperlotti, B.~Goodwine, {Implicit and
  fractional-derivative operators in infinite networks of integer-order
  components}, Chaos, Solitons \& Fractals 114 (2018) 186--192.
\newblock \href {https://doi.org/10.1016/j.chaos.2018.07.003}
  {\path{doi:10.1016/j.chaos.2018.07.003}}.

\bibitem{elwakil2021equivalent}
A.~S. Elwakil, A.~Allagui, C.~Psychalinos, {On the equivalent impedance of
  two-impedance self-similar ladder networks}, IEEE Transactions on Circuits
  and Systems II: Express Briefs 68~(7) (2021) 2685--2689.
\newblock \href {https://doi.org/10.1109/TCSII.2021.3057961}
  {\path{doi:10.1109/TCSII.2021.3057961}}.

\bibitem{elwakil2022generalizing}
A.~S. Elwakil, S.~Kapoulea, C.~Psychalinos, A.~Allagui, {Generalizing the
  Warburg impedance to a Warburg impedance matrix}, AEU-International Journal
  of Electronics and Communications 150 (2022) 154202.
\newblock \href {https://doi.org/10.1016/j.aeue.2022.154202}
  {\path{doi:10.1016/j.aeue.2022.154202}}.

\bibitem{bisquert2002theory}
J.~Bisquert, {Theory of the impedance of electron diffusion and recombination
  in a thin layer}, The Journal of Physical Chemistry B 106~(2) (2002)
  325--333.
\newblock \href {https://doi.org/10.1021/jp011941g}
  {\path{doi:10.1021/jp011941g}}.

\bibitem{king2012percolation}
P.~J. King, T.~M. Higgins, S.~De, N.~Nicoloso, J.~N. Coleman, {Percolation
  effects in supercapacitors with thin, transparent carbon nanotube
  electrodes}, Acs Nano 6~(2) (2012) 1732--1741.
\newblock \href {https://doi.org/10.1021/nn204734t}
  {\path{doi:10.1021/nn204734t}}.

\bibitem{vasilyev2019connections}
O.~A. Vasilyev, A.~A. Kornyshev, S.~Kondrat, {Connections matter: on the
  importance of pore percolation for nanoporous supercapacitors}, ACS Applied
  Energy Materials 2~(8) (2019) 5386--5390.
\newblock \href {https://doi.org/10.1021/acsaem.9b01069}
  {\path{doi:10.1021/acsaem.9b01069}}.

\bibitem{lei2021n}
E.~Lei, J.~Sun, W.~Gan, Z.~Wu, Z.~Xu, L.~Xu, C.~Ma, W.~Li, S.~Liu, {N-doped
  cellulose-based carbon aerogels with a honeycomb-like structure for
  high-performance supercapacitors}, Journal of Energy Storage 38 (2021)
  102414.
\newblock \href {https://doi.org/10.1016/j.est.2021.102414}
  {\path{doi:10.1016/j.est.2021.102414}}.

\bibitem{goodwin2022gelation}
Z.~A. Goodwin, M.~McEldrew, J.~Pedro~de Souza, M.~Z. Bazant, A.~A. Kornyshev,
  {Gelation, clustering, and crowding in the electrical double layer of ionic
  liquids}, The Journal of Chemical Physics 157~(9) (2022).
\newblock \href {https://doi.org/10.1063/5.0097055}
  {\path{doi:10.1063/5.0097055}}.

\bibitem{sugi2002frequency}
M.~Sugi, Y.~Hirano, Y.~Miura, K.~Saito, {Frequency behavior of self-similar
  ladder circuits}, Colloids and Surfaces A: Physicochemical and Engineering
  Aspects 198 (2002) 683--688.
\newblock \href {https://doi.org/10.1016/S0927-7757(01)00988-8}
  {\path{doi:10.1016/S0927-7757(01)00988-8}}.

\bibitem{arbuzov2009three}
A.~A. Arbuzov, R.~R. Nigmatullin, {Three-dimensional fractal models of
  electrochemical processes}, Russian Journal of Electrochemistry 45 (2009)
  1276--1286.
\newblock \href {https://doi.org/10.1134/S1023193509110081}
  {\path{doi:10.1134/S1023193509110081}}.

\bibitem{larminie2003fuel}
J.~Larminie, A.~Dicks, M.~S. McDonald, {Fuel cell systems explained}, Vol.~2,
  J. Wiley Chichester, UK, 2003.
\newblock \href {https://doi.org/10.1002/9781118878330}
  {\path{doi:10.1002/9781118878330}}.

\bibitem{RCsimulator}
V.~G. Malyshkin, {RC simulation program for ngspice.
  \url{http://www.ioffe.ru/LNEPS/malyshkin/RCcircuit_ver2.zip}} (2023).
\newblock
  \href{http://www.ioffe.ru/LNEPS/malyshkin/RCcircuit_ver2.zip}{[link]}.
\newline\urlprefix\url{http://www.ioffe.ru/LNEPS/malyshkin/RCcircuit_ver2.zip}

\bibitem{meurer2017sympy}
A.~Meurer, C.~P. Smith, M.~Paprocki, O.~{\v{C}}ert{\'\i}k, S.~B. Kirpichev,
  M.~Rocklin, A.~Kumar, S.~Ivanov, J.~K. Moore, S.~Singh, et~al., {SymPy:
  symbolic computing in Python}, PeerJ Computer Science 3 (2017) e103.
\newblock \href {https://doi.org/10.7717/peerj-cs.103}
  {\path{doi:10.7717/peerj-cs.103}}.

\bibitem{dutta2015resistive}
S.~Dutta~Roy, {On resistive ladder networks for use in ultra-low frequency
  active-RC filters}, Circuits, Systems, and Signal Processing 34~(11) (2015)
  3661--3670.
\newblock \href {https://doi.org/10.1007/s00034-015-0012-x}
  {\path{doi:10.1007/s00034-015-0012-x}}.

\bibitem{kartci2020history}
A.~Kartci, N.~Herencsar, J.~T. Machado, L.~Brancik, {History and progress of
  fractional-order element passive emulators: A review}, Radioengineering
  29~(2) (2020).
\newblock \href {https://doi.org/10.13164/re.2020.0296}
  {\path{doi:10.13164/re.2020.0296}}.

\end{thebibliography}

\end{document}